\documentclass[prx,twocolumn]{revtex4-2}
\usepackage{amsmath}
\usepackage{amssymb}
\usepackage{graphicx}
\usepackage{braket}
\usepackage{stmaryrd}
\usepackage{hyperref}
\usepackage{color}
\usepackage{dsfont}
\usepackage{bm}

\hypersetup{
    colorlinks,
    citecolor=blue,
    linkcolor=blue,
    urlcolor=blue
}

\begin{document}

\title{
Entropic Trade-off Relations in Stochastic Thermodynamics\\ via Replica Markov Processes
}
\author{Yoshihiko Hasegawa}
\email{hasegawa@biom.t.u-tokyo.ac.jp}
\affiliation{Department of Information and Communication Engineering, Graduate
School of Information Science and Technology, The University of Tokyo,
Tokyo 113-8656, Japan}

\date{\today}
\begin{abstract}

Traditional thermodynamic trade-off relations usually apply to quantities that depend linearly on probability distributions. In contrast, many important information-theoretic measures, such as entropies, are nonlinear and therefore difficult to analyze with existing frameworks.
Motivated by replica methods in quantum information and spin-glass theory, we introduce \emph{replica Markov processes}, i.e., Markovian dynamics of $K$ independent, identical copies, and derive trade-off relations that bound relative moments of replica observables in terms of the dynamical activity. 
By choosing appropriate replica observables, these inequalities translate into bounds on nonlinear quantities of the original single process. 
Focusing on entropic measures of uncertainty, we obtain upper bounds on both the time derivative and the value of the Tsallis entropy for general trajectory-observable distributions.
Moreover, we derive upper bounds on the R\'enyi and Tsallis entropies, where the bounds involve the initial dynamical activity. 
Analogous bounds can be derived for the entropies of the state distribution, which quantify the degree of diffusion. 
In particular, we provide upper bounds on both the time derivative and the value of the Tsallis entropy of state distributions.
Moreover, we show that the R\'enyi and Tsallis entropies of the state distribution are bounded from above by terms involving the local escape rate from the initial node. 
We also illustrate how replicas can be used to study other nonlinear functionals such as extreme-value observables. 
Finally, we extend the construction to continuously monitored open quantum systems governed by Lindblad dynamics, where the bounds are expressed in terms of the quantum dynamical activity. These results provide entropic counterparts to activity-based uncertainty relations and establish a general method for constraining nonlinear information-theoretic quantities in stochastic and quantum thermodynamics.

\end{abstract}
\maketitle

\section{Introduction}

Over the past decade,
thermodynamic trade-off relations have been extensively studied in stochastic and quantum thermodynamics.
These relations express a ``no free lunch'' principle in thermodynamics: faster, more precise operations require greater thermodynamic resources.
In the thermodynamic uncertainty relations \cite{Barato:2015:UncRel,Gingrich:2016:TUP,Garrahan:2017:TUR,Dechant:2018:TUR,Terlizzi:2019:KUR,Hasegawa:2019:CRI,Hasegawa:2019:FTUR,Timpanaro:2019:EFTTUR,Dechant:2020:FRIPNAS,Vo:2020:TURCSLPRE,Koyuk:2020:TUR,Saryal:2019:TUR,Prech:2024:ClockUR}, 
trajectory observables are considered, and lower bounds on their relative variance are derived, where the bounds involve the entropy production and/or the dynamical activity.
Extensions of these uncertainty relations to the quantum domain have been actively studied \cite{Erker:2017:QClockTUR,Brandner:2018:Transport,Carollo:2019:QuantumLDP,Guarnieri:2019:QTURPRR,Liu:2019:QTUR,Hasegawa:2020:QTURPRL,Hasegawa:2020:TUROQS,Rignon-Bret:2021:quantum-precision,Sacchi:2021:BosonicTUR,Vu:2021:QTURPRL,Kalaee:2021:QTURPRE,Moreira:2025:PrecisionBoundsMultipleCurrents,Prech:2025:CoherenceQTUR,Ishida:2024:QTURVerification}.
Related work has also examined how to compute state-dependent expectation values \cite{Nicholson:2020:TIUncRel,GarciaPintos:2022:OSP}.
What these relations have in common is that the quantities being computed are linear in the underlying probabilities.
For example, in thermodynamic uncertainty relations, we are often interested in quantities of the form  
\begin{align}
    \mathbb{E}[f] = \sum_\Gamma \mathcal{P}(\Gamma)\, f(\Gamma),
    \label{eq:Ef_def}
\end{align}
where $\Gamma$ denotes a stochastic trajectory, $\mathcal{P}(\Gamma)$ is the trajectory probability, and $f(\Gamma)$ is an observable, such as a current. 
Equation~\eqref{eq:Ef_def} is linear with respect to the probability $\mathcal{P}(\Gamma)$. 
Typical examples of Eq.~\eqref{eq:Ef_def} include dissipated heat and stochastic displacement.
In contrast, nonlinear functions of the probabilities are generally much harder to compute, although they play a crucial role. 
A prominent example is generalized entropies such as R\'enyi and Tsallis entropies, which measure the degree of uncertainty and the spread of information.
The thermodynamic uncertainty relation expresses uncertainty using the variance of an observable. However, in information theory, uncertainty is often quantified using entropy. When the output is categorical rather than numerical, variance is not a natural measure of uncertainty. Entropy, by contrast, applies directly to categorical outcomes and can be used to quantify uncertainty in this setting.
For example, when quantifying the extent to which a random walker on a network has propagated over the network, variance is not suitable because the states are labels rather than numerical values.
Establishing thermodynamic bounds for these nonlinear functionals remains a challenge because standard frameworks are inherently tied to linear expectations. There is thus a fundamental need for a theoretical bridge that maps nonlinear information measures onto stochastic thermodynamics.

Inspired by replica methods in quantum information and spin-glass theory, we introduce replica Markov process comprising $K$ identical copies of the process and derive bounds on quantities that depend nonlinearly on the underlying probability distributions.
In quantum information and statistical mechanics, we often need to compute the expectation values of nonlinear physical quantities. For example, purity requires the square of the density operator, $\mathrm{Tr}[\rho^2]$.
Although purity serves as a key quantity for distinguishing quantum behavior from classical behavior, its experimental measurement is known to be difficult.
The swap trick \cite{Ekert:2002:DirectEstimations} is a technique to measure purity without performing state tomography. 
The swap trick uses the following identity:
\begin{align}
\mathrm{Tr}[\rho^2] = \mathrm{Tr}\!\left[(\rho \otimes \rho)\,\mathrm{SWAP}\right],
\label{eq:SWAP_trick}
\end{align}
where $\mathrm{SWAP}$ denotes the swap operator. 
The identity given in Eq.~\eqref{eq:SWAP_trick} shows that the purity $\mathrm{Tr}[\rho^2]$ can be obtained by preparing two replicas of $\rho$ and measuring the expectation value of the SWAP operator acting on $\rho \otimes \rho$.
This avoids full state tomography of $\rho$, which would require many measurement shots to reliably estimate $\rho^2$.
The use of replicas has been highly effective in classical shadow tomography \cite{Aaronson:2018:ShadowTomography,Huang:2020:ClassicalShadows,Elben:2023:RandomizedMeasurementToolbox}. 
The use of \textit{hypothetical} replicas for the evaluation of nonlinear quantities has a long history.
In spin glasses \cite{Edwards:1975:SpinGlassTheory,Sherrington:1975:SolvableSpinGlass,Mezard:1986:SpinGlass}, for example, we often need to evaluate the disorder average of the logarithm of the partition function $\mathbb{E}[\ln Z]$, which is difficult to compute directly. The replica method addresses this by introducing $n$ replicas of the system and expressing the logarithm as the limit $\ln Z = \lim_{n \to 0} (Z^{n} - 1)/n$. This technique has been widely used in the analysis of learning theory \cite{Sompolinsky:1990:learning,Opper:1991:BayesClassification,Seung:1992:StatisticalMechanicsLearning,Watkin:1993:StatisticalMechanicsLearning,Nishimori:2001:SpinGlassesInformation} and conformal field theory \cite{Holzhey:1994:EntropyCFT,Calabrese:2004:EntanglementEntropyQFT} .
These examples illustrate the power of using replicas to calculate the expectations of nonlinear quantities.

We introduce a method for deriving trade-off relations for nonlinear quantities of a single Markov process. This is achieved by considering $K$ independent, identical copies (replicas) of the process and applying trade-off relations to suitably chosen replica observables. Through this approach, these nonlinear quantities are expressed as linear expectations within the replica framework. Crucially, replicas can be treated as purely virtual, analogous to the replica trick, allowing us to obtain fundamental constraints on the original single process.
Our work then focuses on entropic measures of uncertainty. We derive upper bounds on both the time derivative and the value of the Tsallis entropy for general trajectory-observable distributions, relating these to dynamical activity [Eqs.~\eqref{eq:dt_HT_traj_upperbound} and \eqref{eq:HT_traj_upperbound}]. Additionally, we establish upper bounds on the R\'enyi and Tsallis entropies, which are constrained by the initial dynamical activity [Eqs.~\eqref{eq:renyi_alpha_bound} and \eqref{eq:tsallis_alpha_bound_traj}]. 
Analogous results are also obtained for state distribution of a Markov process. 
Specifically, we obtain bounds on the time derivative [Eq.~\eqref{eq:dt_HT_state_upperbound}] and the value of the Tsallis entropy for state distributions [Eq.~\eqref{eq:HT_state_upperbound}], where these bounds are determined by the dynamical activity. 
Furthermore, we establish bounds on the R\'enyi and Tsallis entropies of the state distribution, relating them to the initial dynamical activity [Eqs.~\eqref{eq:renyi_alpha_bound_state} and \eqref{eq:tsallis_alpha_bound_state}].
A notable finding is that these initial-activity-controlled bounds depend solely on the local escape rate from the initial node [Eqs.~\eqref{eq:renyi_alpha_bound_state} and \eqref{eq:tsallis_alpha_bound_state}].
Beyond entropy, we demonstrate the utility of replicas for handling other nonlinear functionals, including extreme-value observables [Eqs.~\eqref{eq:maxKUR_def}--\eqref{eq:NM_max_Tsallis_upperbound2}]. Finally, we generalize this construction to continuously monitored open quantum systems governed by Lindblad dynamics, where the resulting bounds are formulated in terms of quantum dynamical activity [Eq.~\eqref{eq:dt_HT_traj_upperbound_quantum}, \eqref{eq:HT_traj_upperbound_quantum}, and \eqref{eq:quantum_Renyi_bound}].
These results provide entropic counterparts to activity-based uncertainty relations. While conventional uncertainty relations typically lower-bound relative variances by the inverse activity, our bounds similarly show that entropy-based uncertainty is fundamentally constrained by dynamical activity.

\begin{figure}
\includegraphics[width=1\linewidth]{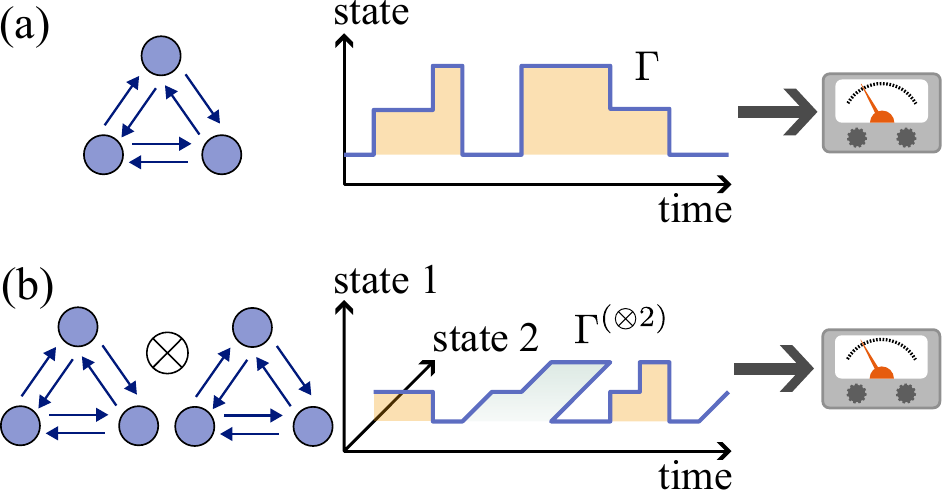}
\caption{
Illustration of Markov processes. 
(a) Single process scenario with $D=3$ states.
This scenario is usually considered in stochastic thermodynamics. 
For a Markov process, we consider stochastic trajectories, which are random sample paths that represent individual realizations of the process.
We are interested in stochastic quantities associated with the trajectories, whose statistics are linear in the trajectory probability. 
(b) $K=2$ replica process scenario with $D^2=9$ states. 
We consider two identical Markov processes. 
The state space is the product space of each Markov process, and the trajectories are defined on the product space. 
We are interested in stochastic quantities of the replica process, which become nonlinear quantities in the original single process. 
}
\label{fig:replica}
\end{figure}

\section{Results}

\subsection{Replica Markov processes}

Consider a classical Markov process with $D$ states, represented by $\mathfrak{B} \equiv \{B_1, B_2, \ldots, B_D\}$ (Fig.~\ref{fig:replica}(a)). 
Let $p_\mu(t)$ be the probability that the system is in state $B_\mu$ at time $t$, and let $W_{\mu^\prime\mu}$ be the transition rate from state $B_\mu$ to state $B_{\mu^\prime}$. 
Here, we assume that the transition rate is time-independent. 
The time evolution is governed by the master equation:
\begin{align}
    \frac{d}{dt}p_{\mu}(t)=\sum_{\mu^{\prime}}W_{\mu\mu^{\prime}}p_{\mu^{\prime}}(t),
    \label{eq:master_eq}
\end{align}
where the diagonal elements are defined by $W_{\mu\mu}=-\sum_{\mu^{\prime}\neq\mu}W_{\mu^{\prime}\mu}$.
In stochastic thermodynamics, the trajectories generated by the master equation [Eq.~\eqref{eq:master_eq}] are fundamental for analyzing the system's stochastic dynamics.
Several quantities, such as state displacement, dissipated heat, and Shannon entropy, are considered at the level of stochastic trajectories. 
We are interested in the dynamics within the time interval $[0,\tau]$ ($\tau > 0$).
Suppose that there are $J$ jump events within the interval.
Let $t_j$ be the time of the $j$-th jump and let 
$X_j\in \mathfrak{B}$ be the state after the $j$-th jump. 
The trajectory $\Gamma$ is then expressed by
\begin{align}
    \Gamma=[(t_{0},X_{0}),(t_{1},X_{1}),\ldots,(t_{J},X_{J})],
    \label{eq:trajectory_single_def}
\end{align}
where $t_0=0$ is the initial time and $X_0$ is the initial state. 
Let $N(\Gamma)$ be an arbitrary real-valued function defined on a trajectory $\Gamma$.
We consider a subclass of the observable $N(\Gamma)$,
 denoted by $N_\circ(\Gamma)$.
The observable $N_\circ(\Gamma)$ satisfies
\begin{align}
    N_{\circ}(\Gamma_{\varnothing})=0,
    \label{eq:zero_condition}
\end{align}
where $\Gamma_\varnothing$ represents a trajectory with no jumps.
An important subclass of $N_\circ(\Gamma)$ is the counting observable, defined as a linear combination of jump counts:
\begin{align}
    N_{\sharp}(\Gamma)\equiv\sum_{\nu,\mu(\nu\ne\mu)}C_{\nu\mu}N_{\nu\mu},
    \label{eq:counting_obs}
\end{align}
where $N_{\nu\mu}$ is the number of jumps from $B_\mu$ to $B_\nu$ within the interval $[0,\tau]$
and $C_{\nu\mu}$ are their associated real weights.
Clearly, $N_{\sharp}(\Gamma)$ satisfies Eq.~\eqref{eq:zero_condition}.
If we denote the sets of functions $N$, $N_\circ$, and $N_{\sharp}$
by $\{N\}$, $\{N_\circ\}$, and $\{N_{\sharp}\}$, respectively,
then the following inclusion holds:
\begin{align}
    \{N_{\sharp}\}\subset\{N_{\circ}\}\subset\{N\}.
    \label{eq:classical_inclusion}
\end{align}

As mentioned in the introduction, we consider replicas of the original process and regard the whole Markov process as a combined process (Fig.~\ref{fig:replica}(b)). 
For simplicity, here we introduce the case of two replicas.
That is, there are two identical processes, with \textit{no} interaction between them. 
It is straightforward to extend this case to scenarios where there are $K$ replica processes. 
The state space is a product of the two state sets.
The set of states in the combined Markov process of two replicas is given by
$\mathfrak{B}^{(\otimes2)}\equiv\{(B_{1}^{(1)},B_{1}^{(2)}),(B_{1}^{(1)},B_{2}^{(2)}),\ldots,(B_{D}^{(1)},B_{D}^{(2)})\}$.
Therefore, the total number of states in the combined process is $|\mathfrak{B}^{(\otimes 2)}| = D^2$. 
From now on, when using a superscript $(k)$, it denotes a quantity associated with the $k$-th process, and when using a superscript $(\otimes K)$, it denotes a quantity for the combined system in which all processes $1,2,\dots,K$ are regarded as a single combined process.
When we omit the superscript, the corresponding variables refer to those in the single process (i.e., $K=1$ case). 

Let $p_{\mu\nu}^{(\otimes 2)}(t)$ be the probability that the combined system is in state $(B_\mu^{(1)},B_{\nu}^{(2)})\in \mathfrak{B}^{(\otimes 2)}$ at time $t$.
Let $W_{\nu\mu}^{(k)}$ be the transition rate from state $B_\mu^{(k)}$ to state $B_\nu^{(k)}$ in the $k$-th process. 
Since we consider identical processes, $W_{\mu^{\prime}\mu}^{(k)}=W_{\mu^{\prime}\mu}$, we keep the superscript for clarity. 
The master equation corresponding to the combined process of the two replicas is given by
\begin{align}
    \frac{d}{dt}p_{\mu\nu}^{(\otimes2)}(t)=\sum_{\mu^{\prime}}W_{\mu\mu^{\prime}}^{(1)}p_{\mu^{\prime}\nu}^{(\otimes2)}(t)+\sum_{\nu^{\prime}}W_{\nu\nu^{\prime}}^{(2)}p_{\mu\nu^{\prime}}^{(\otimes2)}(t).
    \label{eq:master_eq_two}
\end{align}
As in the case of a single process [Eq.~\eqref{eq:master_eq}], we can consider a trajectory in the combined process of the two replicas. 
Suppose that there are $J$ jump events in the trajectory of the combined process within $[0,\tau]$.
Let $t_j^{(\otimes 2)}$ be the time of the $j$-th jump and let 
$X_j^{(\otimes 2)}\in \mathfrak{B}^{(\otimes 2)}$ be the state after the $j$-th jump in the combined process.
Then, the trajectory is expressed as
\begin{align}
    \Gamma^{(\otimes2)}=[(t_{0}^{(\otimes2)},X_{0}^{(\otimes2)}),(t_{1}^{(\otimes2)},X_{1}^{(\otimes2)}),\ldots,(t_{J}^{(\otimes2)},X_{J}^{(\otimes2)})],
    \label{eq:trajectory_two}
\end{align}
where $t_0^{(\otimes 2)}=0$ and $X_{0}^{(\otimes2)}$ is the initial state. 
For instance, suppose that the trajectories of the first and second processes are
\begin{align}
    \Gamma^{(1)}&=[(0,B_{1}^{(1)}),(1,B_{2}^{(1)})],\label{eq:Gamma_one}\\
    \Gamma^{(2)}&=[(0,B_{2}^{(2)}),(2,B_{3}^{(2)}),(3,B_{1}^{(2)})],\label{eq:Gamma_two}
\end{align}
The trajectory in the combined process becomes
\begin{widetext}
\begin{align}
    \Gamma^{(\otimes2)}=\Gamma^{(\otimes2)}(\Gamma^{(1)},\Gamma^{(2)})=[(0,(B_{1}^{(1)},B_{2}^{(2)})),(1,(B_{2}^{(1)},B_{2}^{(2)})),(2,(B_{2}^{(1)},B_{3}^{(2)})),(3,(B_{2}^{(1)},B_{1}^{(2)}))].
    \label{eq:Gamma_combined}
\end{align}
\end{widetext}
where the expression $\Gamma^{(\otimes2)}(\Gamma^{(1)},\Gamma^{(2)})$ emphasizes that $\Gamma^{(\otimes 2)}$ can be constructed given $\Gamma^{(1)}$ and $\Gamma^{(2)}$. 
Here, we have only considered the two-process case. 
However, it is straightforward to generalize the discussion to more than two processes.

\subsection{Replica trade-off relations}

Introducing the combined process with two replicas, 
we will focus on a general scenario in which there are $K$ replicas. 
We now introduce observables associated with trajectories. 
In the same way as in the single process case, we can consider observables $N^{(\otimes K)}(\Gamma^{(\otimes K)})$, $N_\circ^{(\otimes K)}(\Gamma^{(\otimes K)})$, and $N_{\sharp}^{(\otimes K)}(\Gamma^{(\otimes K)})$. 
Specifically, 
The observable $N^{(\otimes K)}$ is an arbitrary function of $\Gamma^{(\otimes K)}$. 
The observable $N_\circ^{(\otimes K)}(\Gamma_{\varnothing}^{(\otimes K)})$ satisfies
\begin{align}
    N_\circ^{(\otimes K)}(\Gamma_{\varnothing}^{(\otimes K)})=0,
    \label{eq:NK_null_condition}
\end{align}
where $\Gamma_\varnothing^{(\otimes K)}$ is a trajectory in the combined process with no jumps. 
$N_{\sharp}^{(\otimes K)}$ is the counting observable, which yields the weighted sum of the number of jumps within the interval $[0,\tau]$ in the $K$-replica process.

Let $\mathbb{E}[X]$ denote the expectation value of a random variable $X$. 
Since $\Gamma^{(\otimes K)}$ is a random variable, the observable $N^{(\otimes K)}(\Gamma^{(\otimes K)})$ is also a random variable. 
From the uncertainty relation in Ref.~\cite{Terlizzi:2019:KUR}, the following relation holds in the Markov process with $K$ replicas:
\begin{align}
    \frac{\mathrm{Var}[N^{(\otimes K)}]}{\tau^{2}\left(\partial_{\tau}\mathbb{E}[N^{(\otimes K)}]\right)^{2}}\ge\frac{1}{K\mathcal{A}(\tau)}.
    \label{eq:transient_KUR}
\end{align}
Here, $\mathcal{A}(\tau)$ is the time-integrated dynamical activity:
\begin{align}
    \mathcal{A}(\tau) = \int_0^\tau \mathfrak{a}(t)dt,
    \label{eq:Atau_def}
\end{align}
where $\mathfrak{a}(t)$ is the dynamical activity at time $t$:
\begin{align}
    \mathfrak{a}(t)\equiv\sum_{\mu^{\prime},\mu(\mu\ne\mu^{\prime})}W_{\mu^{\prime}\mu}p_{\mu}(t).
    \label{eq:dynamical_activity_def}
\end{align}
Note that the dynamical activity $\mathfrak{a}(t)$ in Eq.~\eqref{eq:dynamical_activity_def} is a quantity of a single process [Eq.~\eqref{eq:master_eq}]; the definition is the same as that in conventional Markov process settings. 
Dynamical activity was originally proposed as an order parameter in the study of glass transitions \cite{Biroli:2013:GlassTransition}.  
The dynamical activity plays a central cost term in thermodynamic trade-off relations \cite{Garrahan:2017:TUR,Shiraishi:2018:SpeedLimit,Terlizzi:2019:KUR,Hasegawa:2023:BulkBoundaryBoundNC}. 
Since the replica process comprises $K$ independent processes, the dynamical activity of the combined process is $K\mathcal{A}(\tau)$, which directly yields Eq.~\eqref{eq:transient_KUR}.

Moreover, by employing another trade-off relation in Ref.~\cite{Hasegawa:2024:ConcentrationIneqPRL},
we can obtain another relation for the replica process.
Recall that $N_\circ(\Gamma)$ is the observable that satisfies the condition given in Eq.~\eqref{eq:zero_condition}. 
We obtain (see Appendix~\ref{sec:classical_nojump_KUR} for the derivation):
\begin{align}
    \frac{\mathbb{E}\left[|N_{\circ}^{(\otimes K)}|^{s}\right]^{r/(s-r)}}{\mathbb{E}\left[|N_{\circ}^{(\otimes K)}|^{r}\right]^{s/(s-r)}}\geq\frac{1}{1-e^{-\tau K\mathfrak{a}(t=0)}},
    \label{eq:main_result1}
\end{align}
where $0<r<s$.
For $r = 1$ and $s = 2$, Eq.~\eqref{eq:main_result1} reduces to the bound for the relative variance:
\begin{align}
    \frac{\mathrm{Var}\left[N_{\circ}^{(\otimes K)}\right]}{\mathbb{E}\left[N_{\circ}^{(\otimes K)}\right]^{2}}\ge\frac{1}{e^{K\tau\mathfrak{a}(t=0)}-1}.
    \label{eq:transient_nojump_KUR}
\end{align}
Equations~\eqref{eq:transient_KUR} and \eqref{eq:transient_nojump_KUR} have similar forms, but there are several differences.
First, the left-hand side of Eq.~\eqref{eq:transient_KUR} is not written in the form of a relative variance, and the denominator on the left-hand side is given by a derivative of the expectation.
By contrast, the left-hand side of Eq.~\eqref{eq:transient_nojump_KUR} has the usual relative-variance form.
In addition, the right-hand side of Eq.~\eqref{eq:transient_KUR} is the reciprocal of the time-integrated dynamical activity $\mathcal{A}(\tau)$, while on the right-hand side of Eq.~\eqref{eq:transient_nojump_KUR} the dynamical activity appears in the exponent, and only the dynamical activity $\mathfrak{a}(0)$, which is evaluated at $t=0$.

\subsection{Entropic measures}

While the right-hand sides of Eqs.~\eqref{eq:transient_KUR} and \eqref{eq:main_result1} are written entirely in terms of quantities for a single process, the left-hand side contains the observable for the combined process of $K$ replicas. 
Note that Eqs.~\eqref{eq:transient_KUR} and \eqref{eq:main_result1} are intermediate relations that will be used to obtain trade-off relations in the single process. 
In other words, although we consider
the combined process of replicas $K$, 
this is purely for mathematical purposes, and the final results that will be shown later are relations for the single process. 
This is the same idea as in methods such as multi-copy measurements and the replica trick.
As shown in the introduction, in two-copy measurements,
we use $\rho \otimes \rho$ but the quantity obtained from using these copies is $\mathrm{Tr}[\rho^2]$, which is a quantity referring to the case of a single-copy.

\begin{figure*}
\centering
\includegraphics[width=0.9\linewidth]{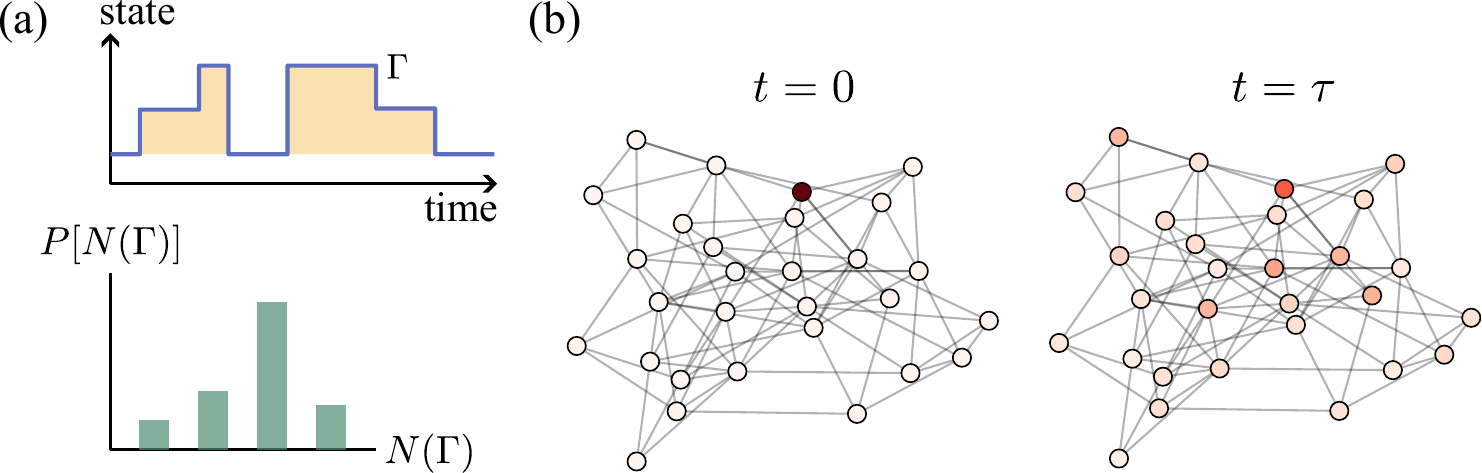}
\caption{
Illustration of two scenarios considered in trade-off relations: (a) the trajectory observable and (b) the state distribution.
(a) Trajectory observable bound in a Markov process. 
For the Tsallis entropy bound for trajectory observables,
we consider a trajectory $\Gamma$ and its associated observable $N(\Gamma)$. 
The entropies are considered for the trajectory observable, and the bounds on the time-derivative and their values are obtained. 
(b) State distribution bound in the diffusion on a network. 
In this scenario, 
a random walker starts from a node in the network, which is represented by dark red. 
At time $t=\tau$, the random walker diffuses over the network, and 
the probability $p_\mu(\tau)$ of the walker position is expressed by red, where darker color corresponds to higher probability. 
The entropies are considered for the state distribution $p_\mu(\tau)$, and the bounds on the time-derivative and their values are obtained. 
}
\label{fig:uncertainty_propagation}
\end{figure*}

By considering specific observables for $N^{(\otimes K)}$ and $N_\circ^{(\otimes K)}$ in Eqs.~\eqref{eq:transient_KUR} and \eqref{eq:main_result1}, we derive trade-off relations for a single process. In particular, we focus on one-parameter generalizations of entropy that quantify different aspects of uncertainty in an arbitrary probability distribution $\{p_n\}$.
One widely used one-parameter generalization is the Tsallis entropy \cite{Tsallis:1988:BoltzmannGibbsGeneralization,Tsallis:2009:Nonextensive}, defined by
\begin{align}
H_{q}^{\mathrm{T}}[p_{n}]\equiv \frac{1-\sum_{n}p_{n}^{q}}{q-1},
\label{eq:Tsallis_def}
\end{align}
where $q$ is the entropic index. Varying $q$ continuously interpolates between several standard entropic quantities: in the limit $q\to 0$, $H_q^{\mathrm T}$ approaches $\mathfrak{W}-1$, where $\mathfrak{W}$ is the number of outcomes with nonzero probability; as $q\to 1$, it reduces to the Shannon entropy; for $q=2$, it yields $H_2^{\mathrm T}=1-\sum_n p_n^2$, i.e., the complement of the collision probability; and in the limit $q\to\infty$, it tends to $0$ for any non-degenerate distribution. For $q>1$, the Tsallis entropy satisfies
\begin{align}
0\le H_{q}^{\mathrm{T}}<\frac{1}{q-1}\;\;(q>1).
\label{eq:Tsallis_range}
\end{align}
Another important one-parameter generalization is the R\'enyi entropy \cite{Renyi:1961:entropy} of order $\alpha$ $(0<\alpha<\infty)$, defined as
\begin{align}
H_{\alpha}^{\mathrm{R}}[p_{n}]\equiv\frac{1}{1-\alpha}\ln\left(\sum_{n}p_{n}^{\alpha}\right).
\label{eq:Renyi_ent_def}
\end{align}
By varying $\alpha$ continuously, we can interpolate between several standard entropy measures: in the limit $\alpha\to 0$, $H_\alpha^{\mathrm R}$ approaches the max-entropy, which depends only on the number of outcomes with nonzero probability; as $\alpha\to 1$, it reduces to the Shannon entropy; for $\alpha=2$, it becomes the collision entropy; and in the limit $\alpha\to\infty$, it converges to the min-entropy determined solely by the largest probability.
The two entropies are related via
\begin{align}
    H_{\alpha}^{\mathrm{R}}&=\frac{1}{1-\alpha}\ln\left[1+(1-\alpha)H_{\alpha}^{\mathrm{T}}\right],
    \label{eq:Renyi_and_Tsallis}\\
    H_{q}^{\mathrm{T}}&=\frac{\exp\left((1-q)H_{q}^{\mathrm{R}}\right)-1}{1-q}.
    \label{eq:Tsallis_and_Renyi}
\end{align}
The Tsallis entropy is mainly applied to non-extensive statistical mechanics and the study of complex systems, whereas the R\'enyi entropy is widely used in statistics and quantum information theory.

\subsection{Entropic bound on trajectory observable}

We consider the uncertainty in trajectory observables. 
Trajectory observables play a fundamental role in uncertainty relations in stochastic thermodynamics (Fig.~\ref{fig:uncertainty_propagation}(a)).
Consider a general observable $N(\Gamma)$. 
Using Eq.~\eqref{eq:transient_KUR}, we obtain the upper bound on the time derivative of the Tsallis entropy (see Appendix~\ref{sec:derivation_of_timederivative} for the derivation):
\begin{align}
    \left|\frac{\partial}{\partial\tau}H_{q}^{\mathrm{T}}\left[P\left(N\right)\right]\right|\leq\frac{\sqrt{q\mathcal{A}(\tau)}}{2\tau(q-1)},
    \label{eq:dt_HT_traj_upperbound}
\end{align}
where $q = 2, 3, \cdots$. 
The bound in Eq.~\eqref{eq:dt_HT_traj_upperbound} provides the upper bound on the change in the Tsallis entropy. 
Equation~\eqref{eq:dt_HT_traj_upperbound} shows that the change in entropy is larger for larger $\mathcal{A}(\tau)$. 
Integrating in Eq.~\eqref{eq:dt_HT_traj_upperbound}, we can derive the upper bound on the Tsallis entropy with respect to the observable $N_\circ$ (see Appendix~\ref{sec:derivation_of_timederivative} for the derivation):
\begin{align}
    H_{q}^{\mathrm{T}}[P(N_{\circ})]\le\frac{1}{q-1}\sin\left(\frac{1}{2}\int_{0}^{\tau}\frac{\sqrt{q\mathcal{A}(t)}}{t}dt\right)^{2},
    \label{eq:HT_traj_upperbound}
\end{align}
where $q = 2, 3,\cdots$.
Note that 
in Eq.~\eqref{eq:HT_traj_upperbound}, the following condition should be satisfied:
\begin{align}
    0\le\frac{1}{2}\int_{0}^{\tau}\frac{\sqrt{q\mathcal{A}(t)}}{t}dt\le\frac{\pi}{2}.
    \label{eq:At_range_condition}
\end{align}
Outside the range of Eq.~\eqref{eq:At_range_condition}, the bound of Eq.~\eqref{eq:HT_traj_upperbound} provides a trivial upper bound $1/(q-1)$, which directly follows from Eq.~\eqref{eq:Tsallis_range}. 
On the right-hand side of Eq.~\eqref{eq:HT_traj_upperbound}, the integral $\int_0^\tau dt\, \sqrt{\mathcal{A}(t)}/t$ often appears in thermodynamic trade-off relations \cite{Hasegawa:2023:BulkBoundaryBoundNC}.

We have demonstrated the consequence of considering Eq.~\eqref{eq:transient_KUR} in the replica scenario. 
We can derive another trade-off relation from Eq.~\eqref{eq:main_result1}. 
From Eq.~\eqref{eq:main_result1}, we obtain another trade-off relation (see Appendix~\ref{sec:derivation_entropic_bound} for the derivation):
\begin{align}
    H_{\alpha}^\mathrm{R}[P(N_\circ)]\le\frac{\alpha\tau}{\alpha-1}\mathfrak{a}(t=0),
    \label{eq:renyi_alpha_bound}
\end{align}
where $\alpha=2,3,\cdots$. 
The left-hand side of Eq.~\eqref{eq:renyi_alpha_bound} is the R\'enyi entropy of the observable $N_\circ$, while the right-hand side depends on the dynamical activity and the duration $\tau$.
For example, consider dynamics with no jumps at all, or the case $\tau = 0$. Since there are no jumps, the distribution of $N_\circ$ is $P(N_\circ = 0) = 1$, and the entropy of $N_\circ(\Gamma)$ is zero. In this case, the right-hand side also vanishes, so the result is consistent.
On the other hand, for dynamics with a very large number of jumps, the distribution acquires weight at many different values of $N_\circ$, and the entropy increases. In fact, the right-hand side increases with both the dynamical activity and duration $\tau$, which is consistent with our intuition.
Equation~\eqref{eq:renyi_alpha_bound} is equivalently written as the bound for the Tsallis entropy via Eq.~\eqref{eq:Tsallis_and_Renyi}:
\begin{align}
    H_{q}^{\mathrm{T}}[P(N_{\circ})]\le\frac{1-e^{-q\tau\mathfrak{a}(t=0)}}{q-1}.
    \label{eq:tsallis_alpha_bound_traj}
\end{align}
Both left-hand sides of Eqs.~\eqref{eq:HT_traj_upperbound} and \eqref{eq:tsallis_alpha_bound_traj} are Tsallis entropies, and the right-hand sides involve the dynamical activity. In a general transient setting, it is unclear which of the two is tighter; however, in the steady state, Eq.~\eqref{eq:tsallis_alpha_bound_traj} provides a tighter bound.

Traditionally, uncertainty in trajectory observables has been characterized using the variance $\mathrm{Var}[N_{\sharp}]$, whose relative value is known to be bounded from below by the inverse of the dynamical activity. 
For a steady-state Markov process and the counting observable $N_{\sharp}(\Gamma)$ [Eq.~\eqref{eq:counting_obs}], it is known that the following relation holds \cite{Garrahan:2017:TUR,Terlizzi:2019:KUR}:
\begin{align}
    \frac{\mathrm{Var}[N_{\sharp}]}{\mathbb{E}[N_{\sharp}]^{2}}\ge\frac{1}{\mathfrak{a}_{\mathrm{ss}}\tau},
    \label{eq:KUR_def}
\end{align}
where $\mathfrak{a}_\mathrm{ss}$ is the steady-state dynamical activity. 
Moreover, under non-steady-state conditions and for the observable $N_\circ(\Gamma)$ satisfying Eq.~\eqref{eq:NK_null_condition}, the following relation is known to hold \cite{Hasegawa:2024:ConcentrationIneqPRL,Nishiyama:2024:NonHermiteQSLPRA}:
\begin{align}
    \frac{\mathrm{Var}[N_{\circ}]}{\mathbb{E}[N_{\circ}]^{2}}\ge\frac{1}{e^{\mathfrak{a}(t=0)\tau}-1},
    \label{eq:KUR_transient_def}
\end{align}
which follows directly from Eq.~\eqref{eq:main_result1} with $K=1$. 
Our results show that this fundamental constraint extends beyond relative variance: the uncertainty measure based on Tsallis (or R\'enyi) entropy is also limited by the dynamical activity.  
However, there is an important difference between Eq.~\eqref{eq:tsallis_alpha_bound_traj} [or Eq.~\eqref{eq:HT_traj_upperbound}] and Eq.~\eqref{eq:KUR_def}.  
When the variance $\mathrm{Var}[N_\sharp]$ decreases, the uncertainty in $N_\sharp$ becomes smaller; thus, Eq.~\eqref{eq:KUR_def} provides a lower bound on the uncertainty in $N_\sharp$.  
By contrast, Eq.~\eqref{eq:tsallis_alpha_bound_traj} [or Eq.~\eqref{eq:HT_traj_upperbound}] yields an upper bound on the uncertainty in the trajectory observable.  
Therefore, these two relations provide complementary lower and upper bounds on the uncertainty of the trajectory observable, each expressed in terms of different quantities.
The Tsallis entropy reduces to the Shannon entropy for $q \to 1$; however, the bound in Eq.~\eqref{eq:tsallis_alpha_bound_traj} is not defined for $q = 1$. 
Regarding the Shannon entropy for trajectory observables,
Ref.~\cite{Hasegawa:2025:EntropicURPRE} derived a lower bound for the Shannon entropy using the entropy production.
The bound in Eq.~\eqref{eq:tsallis_alpha_bound_traj} provides an upper bound on the Tsallis entropy, and this bound involves the dynamical activity.

\subsection{Entropic bound on state distribution}

So far, we have been concerned with entropic trade-off relations for the trajectory observables.
Next, we derive entropic trade-off relations for the state probability distribution $p_\mu(t)$. 
The entropy of the state distribution can quantify the extent of diffusion in a network. 
Suppose that a random walker moves on a network whose dynamics is governed by a continuous-time Markov process [Eq.~\eqref{eq:master_eq}], which has been extensively investigated in the literature \cite{Lovasz:1996:RandomWalksGraphs,Noh:2004:RandomWalksComplexNetworks,Masuda:2017:RandomWalksReview}. 
It has many applications in web search engines \cite{Brin:1998:LargeScaleSearchEngine}, community detection \cite{Pons:2005:CommunitiesRandomWalks}, embedding algorithms \cite{Perozzi:2014:DeepWalk}, and computational biology \cite{Kohler:2008:InteractomePrioritization}, to name but a few. 
Here, our objective is to quantify the extent of diffusion in the network (Fig.~\ref{fig:uncertainty_propagation}(b)).
Typically, the mean squared displacement, mixing time, or the spectral gap of the generator is used \cite{Lovasz:1996:RandomWalksGraphs,Masuda:2017:RandomWalksReview}.
Entropy is another metric used to quantify diffusion in a network.  
The diversity entropy \cite{Travencolo:2008:Accessibility,Viana:2012:EffectiveNodes} is defined as the entropy of the probability distribution of the position of the walker and measures how widely a random walker can spread over the network.  
The entropy rate \cite{Gomez-Gardenes:2008:EntropyDiffusion} quantifies how quickly the uncertainty increases as the random walk evolves.

The degree of diffusion can be quantified by the Tsallis entropy (or R\'enyi entropy) $H_{q}^{\mathrm{T}}[p_{\mu}(\tau)]$ (or $H_\alpha^\mathrm{R}[p_\mu(\tau)]$).
When this entropy is zero, it indicates that no propagation has occurred in the network, and as the degree of propagation increases, the value of this entropy increases.
From Eq.~\eqref{eq:dt_HT_traj_upperbound}, we obtain
\begin{align}
    \left|\frac{\partial}{\partial\tau}H_{q}^{\mathrm{T}}\left[p_{\mu}\left(\tau\right)\right]\right|\leq\frac{\sqrt{q\mathcal{A}(\tau)}}{2\tau(q-1)},
    \label{eq:dt_HT_state_upperbound}
\end{align}
where $q = 2, 3, \cdots$. 
The derivation of Eq.~\eqref{eq:dt_HT_state_upperbound} is shown in Appendix~\ref{sec:derivation_of_timederivative_state}. 
As expected, larger dynamical activity allows for faster diffusion, which is quantified by the change in the Tsallis entropy. 
By integrating Eq.~\eqref{eq:dt_HT_state_upperbound}, we obtain an upper bound on the difference in the Tsallis entropy of the state distribution:
\begin{align}
    \left|\Delta H_{q}^{\mathrm{T}}[p_{\mu}]\right|\leq\frac{\sqrt{q}}{2(q-1)}\int_{0}^{\tau}\frac{\sqrt{\mathcal{A}(t)}}{t}dt,
    \label{eq:dt_HT_state_upperbound_integration}
\end{align}
where $q = 2, 3,\cdots$ and $\Delta H_{q}^{\mathrm{T}}[p_{\mu}]\equiv H_{q}^{\mathrm{T}}[p_{\mu}(\tau)]-H_{q}^{\mathrm{T}}[p_{\mu}(0)]$. 
A similar relation was obtained for $q=2$ in Ref.~\cite{Nishiyama:2025:TemporalFisher}. 
Equation~\eqref{eq:dt_HT_state_upperbound_integration} describes the difference in the Tsallis entropy at $t=0$ and $t=\tau$.
In the Landauer principle \cite{Landauer:1961:LP,Bennett:2003:Landauer}, the following relation holds:
\begin{align}
    \Delta H_{1}^{\mathrm{T}}[p_{\mu}]\ge - \beta\Delta Q(\tau),
    \label{eq:Landauer_bound}
\end{align}
where $\beta$ is the inverse temperature and $H_1^{\mathrm{T}}[p_\mu]\equiv\lim _{q \rightarrow 1} H_q^{\mathrm{T}}[p_\mu]=-\sum_\mu p_\mu \ln p_\mu$ corresponds to the Shannon entropy. 
In Eq.~\eqref{eq:Landauer_bound}, the change in the Shannon entropy $H_1^{\mathrm{T}}\left[p_\mu(t)\right]$ from the initial distribution $p_\mu(0)$ to the distribution at time $\tau$, $p_\mu(\tau)$, is constrained by the heat exchanged with a thermal reservoir. With the sign convention that $\Delta Q(\tau)$ is the heat absorbed by the reservoir up to time $\tau$, this inequality
relates the information change to the thermodynamic cost. 
In contrast, Eq.~\eqref{eq:dt_HT_state_upperbound_integration} provides an upper bound on the change in Tsallis entropy, where the upper bound comprises the dynamical activity. 

The bounds in Eqs.~\eqref{eq:dt_HT_state_upperbound} and \eqref{eq:dt_HT_state_upperbound_integration} hold for an arbitrary initial state $p_\mu(0)$. 
When the initial state is $B_{\mu_0}$ (i.e., $p_{\mu_0}(0)=1$), the tighter relation holds (see Appendix~\ref{sec:derivation_of_timederivative_state} for the derivation):
\begin{align}
    H_{q}^{\mathrm{T}}[p_{\mu}(\tau\mid\mu_{0})]\le\frac{1}{q-1}\sin\left(\frac{1}{2}\int_{0}^{\tau}\frac{\sqrt{q\mathcal{A}(t)}}{t}dt\right)^{2},
    \label{eq:HT_state_upperbound}
\end{align}
where $q = 2, 3,\cdots$, and the condition given in Eq.~\eqref{eq:At_range_condition} should be satisfied.

As in the trajectory observable case, we can derive another trade-off relation from Eq.~\eqref{eq:main_result1}. 
Using Eq.~\eqref{eq:main_result1}, the following relation holds for the R\'enyi entropy (see Appendix~\ref{sec:derivation_entropic_bound}):
\begin{align}
    H_{\alpha}^{\mathrm{R}}[p_{\mu}(\tau\mid\mu_{0})]\le\frac{\alpha\tau}{\alpha-1}\sum_{\mu(\mu\ne\mu_{0})}W_{\mu\mu_{0}},
    \label{eq:renyi_alpha_bound_state}
\end{align}
where $\alpha=2,3,\cdots$. 
From Eq.~\eqref{eq:Tsallis_and_Renyi},
Eq.~\eqref{eq:renyi_alpha_bound_state} is equivalently written as the bound for the Tsallis entropy:
\begin{align}
    H_{q}^{\mathrm{T}}[p_{\mu}(\tau\mid\mu_{0})]\le\frac{1-e^{-q\tau\sum_{\mu(\mu\ne\mu_{0})}W_{\mu\mu_{0}}}}{q-1},
    \label{eq:tsallis_alpha_bound_state}
\end{align}
where $q=2,3,\cdots$. 
The left-hand sides of Eqs.~\eqref{eq:HT_state_upperbound} and \eqref{eq:tsallis_alpha_bound_state} are the same, and, in both equations, the upper bound of the Tsallis entropy at time $\tau$ is given by the dynamical activity; a larger dynamical activity allows for a larger Tsallis entropy.  
On the other hand, while Eq.~\eqref{eq:HT_state_upperbound} requires information beyond the neighborhood of the initial node, Eq.~\eqref{eq:tsallis_alpha_bound_state} does not.
Specifically, in Eq.~\eqref{eq:tsallis_alpha_bound_state}, it is bounded from above by $\tau$ and the local escape rate from the initial node $B_{\mu_0}$, which is the sum of the transition rates from the initial node $B_{\mu_0}$ to the nodes that can be reached in a single jump (note that $\sum_{\mu(\mu\ne\mu_{0})}W_{\mu\mu_{0}}=\mathfrak{a}(t=0)$ corresponds to the dynamical activity at the initial time). 
Therefore, the right-hand side of Eq.~\eqref{eq:tsallis_alpha_bound_state} depends only on the time $\tau$ and the local escape rate from $B_{\mu_0}$ (see Fig.~\ref{fig:local_escape_rate}). 
In other words, regardless of how large the overall network is, we can compute the upper bound of the entropy using only information about the initial node and the transition rates to nodes that are reachable in a single hop.

\subsection{Bound on extreme value}

By using replicas, we can consider more general observables.
As the next example, we derive a trade-off relation for the extreme value of a stochastic process.
Extreme-value statistics studies the behavior of the largest (and similarly, the smallest) value in a collection of random variables $\{X_k\}$; that is, it focuses on the random variable $\max_k X_k$ (and analogously $\min_k X_k$).
Consider the observables $N^{M,\max}$ and $N_\circ^{M,\max}$, which are the maximum of $N^{(k)}(\Gamma^{(k)})$ and $N_\circ^{(k)}(\Gamma^{(k)})$, respectively:
\begin{align}
    N^{M\max}(\Gamma^{(\otimes M)})&\equiv\max_{1\le k\le M}N^{(k)}(\Gamma^{(k)}),\label{eq:NKa_max_def}\\N_{\circ}^{M\max}(\Gamma^{(\otimes M)})&\equiv\max_{1\le k\le M}N_{\circ}^{(k)}(\Gamma^{(k)}).
    \label{eq:NK_max_def}
\end{align}
When the state of the Markov process describes the position,
$N^{M\max}$ and $N^{M\max}_{\circ}$ express the maximum position at time $t=\tau$ in $M$ processes. 
In reliability engineering, the system load can be modeled using a Markov process \cite{Kharoufeh:2003:WearProcessMarkovianEnvironment,Deulkar:2020:StorageSizing}.
In this case, the maximum value of the stochastic process represents the maximum load of the parallel system.
Since $N^{(\otimes K)}$ in Eq.~\eqref{eq:transient_KUR} allows for an arbitrary observable, the following relation holds:
\begin{align}
    \frac{\mathrm{Var}[N^{M\max}]}{\tau^{2}\left(\partial_{\tau}\mathbb{E}[N^{M\max}]\right)^{2}}\ge\frac{1}{M\mathcal{A}(\tau)}.
    \label{eq:transient_KUR_Mmax}
\end{align}
Moreover, since $N_\circ^{M\max}$ satisfies Eq.~\eqref{eq:zero_condition},
using Eq.~\eqref{eq:transient_nojump_KUR}, we obtain
\begin{align}
    \frac{\mathrm{Var}[N_{\circ}^{M\max}]}{\mathbb{E}[N_{\circ}^{M\max}]^{2}}\ge\frac{1}{e^{M\mathfrak{a}(t=0)\tau}-1}.
    \label{eq:maxKUR_def}
\end{align}
Since the dynamical activity is multiplied by $M$, the relative variance of the maximum observable $N_\circ^{M\max}$ can be smaller than that of a single process. 
Usually, when we take the max operation, the mean becomes larger and the variance becomes smaller than those of the original distribution.
For example, consider two independent random variables, each following a standard normal distribution. If we take the maximum of these two variables, the resulting variable will have a mean of $1/\sqrt{\pi}$ and a variance of $1-1/\pi$. Its mean is larger, and its variance is smaller, when compared to the original standard normal distribution.

Replicas can also be used to establish bounds on the entropy of $N_\circ^{M\max}$. We define $K=qM$ replicas, where $q$ is the entropic order and $M$ is the number of observations used to compute the maximum. We then consider the Tsallis entropy of the probability distribution of $N^{M\max}_\circ$. Applying the same technique as outlined in Eqs.~\eqref{eq:dt_HT_traj_upperbound}, \eqref{eq:HT_traj_upperbound}, and \eqref{eq:tsallis_alpha_bound_traj}, we derive the following bounds:
\begin{align}
    \left|\frac{\partial}{\partial\tau}H_{q}^{\mathrm{T}}[P(N^{M\max})]\right|\leq\frac{\sqrt{qM\mathcal{A}(\tau)}}{2\tau(q-1)},
    \label{eq:NM_max_dtau_Tsallis}
\end{align}
\begin{align}
    H_{q}^{\mathrm{T}}[P(N_{\circ}^{M\max})]\le\frac{1}{q-1}\sin\left(\frac{1}{2}\int_{0}^{\tau}\frac{\sqrt{qM\mathcal{A}(t)}}{t}dt\right)^{2},
    \label{eq:NM_max_Tsallis_upperbound1}
\end{align}
and 
\begin{align}
    H_{q}^{\mathrm{T}}[P(N_{\circ}^{M\max})]\le\frac{1-e^{-qM\tau\mathfrak{a}(t=0)}}{q-1},
    \label{eq:NM_max_Tsallis_upperbound2}
\end{align}
where $q=2,3,\cdots$. 
Again, $0\le(1/2)\int_{0}^{\tau}\sqrt{qM\mathcal{A}(t)}/t\,dt\le\pi/2$ should be satisfied in Eq.~\eqref{eq:NM_max_Tsallis_upperbound1}.

\subsection{Quantum replica Markov process}

The bounds given in Eqs.~\eqref{eq:transient_KUR} and \eqref{eq:main_result1} can be generalized to quantum scenarios using the continuous measurement formalism. 

Let $\rho(t)$ denote the density operator of the system at time $t$.  
We assume that its dynamics are governed by a Lindblad master equation:
\begin{align}
    \dot{\rho}(t) = \mathcal{L}\rho(t),
    \label{eq:Linadlad_eq}
\end{align}
where the Lindblad generator $\mathcal{L}$ is given by
\begin{align}
    \mathcal{L}\rho\equiv-i[H,\rho]+\sum_{m=1}^{\mathfrak{N}}\mathcal{D}[L_{m}]\rho.
    \label{eq:Lindblad_op}
\end{align}
Here, $H$ is the system Hamiltonian, $L_m$ is the $m$-th jump operator, $\mathfrak{N}$ is the number of jump channels, and
\begin{align}
    \mathcal{D}[L]\rho\equiv L\rho L^{\dagger}-\frac{1}{2}\{L^{\dagger}L,\rho\},
    \label{eq:Lindblad_dissipator}
\end{align}
is the corresponding dissipator.
The jump operators $L_m$ describe dissipative processes such as energy relaxation, dephasing, and decoherence. Each term represents a distinct decay or noise channel through which the system interacts with its environment. The term $-i[H,\rho]$ in Eq.~\eqref{eq:Lindblad_op} generates the coherent part of the evolution, while the dissipators $\mathcal{D}[L_m]$ represent irreversible, non-unitary dynamics induced by this coupling.
The Lindblad form of $\mathcal{L}$ guarantees that the time evolution is completely positive and trace preserving (CPTP). Equation~\eqref{eq:Linadlad_eq} provides a general description of Markovian open quantum system dynamics.
When the dynamics are restricted to the energy eigenbasis, that is, when the initial state and the jump operators do not include off-diagonal contributions with respect to the energy eigenbasis, the dynamics reduce to the classical Markov process expressed by Eq.~\eqref{eq:master_eq}.

We now consider continuous measurement in the Lindblad framework, corresponding to continuous monitoring of the environment coupled to the system (see Ref.~\cite{Landi:2023:CurFlucReviewPRXQ}). In this setting, the system's stochastic evolution is conditioned on the measurement outcomes registered in the environment, and individual realizations of the dynamics are described by quantum trajectories. The measurement record consists of the types of jump events and their time stamps, which represent the full history of detected quanta (such as emitted photons or transferred particles) over the time interval. 
Suppose that within the interval $[0,\tau]$ there are $J$ jump events, and let $m_j \in \{1,\dots,\mathfrak{N}\}$ denote the type of the $j$-th jump occurring at time $t_j$. 
The complete measurement record is 
\begin{align}
    \Gamma_{Q}\equiv\big[(t_{1},m_{1}),(t_{2},m_{2}),\ldots,(t_{J},m_{J})\big],
    \label{eq:Gamma_q_def}
\end{align}
where the subscript $``Q"$ is used to emphasize that $\Gamma_Q$ is a quantum trajectory. 
In what follows, variables and functions with the subscript $Q$ denote those in the quantum system.
The sequence $\Gamma_Q$ characterizes one realization of the monitored dynamics and is the basic object from which we construct trajectory-dependent observables.
We first introduce the most general class of trajectory observables.
Let $N_{Q}(\Gamma_Q)$ be an arbitrary real-valued function defined on $\Gamma_Q$. We denote the set of all such functions by $\{N_{Q}\}$.
Next, we impose a natural constraint related to the no-jump trajectory.
Let $\Gamma_{Q\varnothing}$ be a quantum trajectory in which there is no jump within $[0,\tau]$.
Analogous to the condition of Eq.~\eqref{eq:NK_null_condition}, we define $N_{Q\circ}(\Gamma_Q)$ as an observable satisfying
\begin{align}
    N_{Q\circ}(\Gamma_{Q\varnothing})=0,
    \label{eq:Nqc_condition}
\end{align}
and denote the set of such observables by $\{N_{Q\circ}\}$.
Finally, we specify a particularly important subclass that depends only on jump counts with fixed weights.
We define the counting-type trajectory observable $N_{Q\sharp}(\Gamma_Q)$ by
\begin{align}
    N_{Q\sharp}(\Gamma_{Q})\equiv \sum_{m}C_{m}N_{m},
    \label{eq:quantum_counting_def}
\end{align}
where $N_m$ is the number of $m$-th jumps within $[0,\tau]$ and $C_m$ denotes the corresponding real weight. We denote the set of such observables by $\{N_{Q\sharp}\}$.  
By construction, $N_{Q\sharp}$ is included in $N_{Q\circ}$ (in particular, it vanishes on $\Gamma_{Q\varnothing}$).
With these definitions, the inclusion relation is
\begin{align}
    \{N_{Q\sharp}\}\subset\{N_{Q\circ}\}\subset\{N_{Q}\}.
    \label{eq:quantum_inclusion}
\end{align}

Analogous to the classical case, we consider $K$ identical replicas of the quantum Markov process.  
Specifically, $\Gamma_{Q}^{(\otimes K)}$ is constructed from the set $\left(\Gamma_{Q}^{(1)},\Gamma_{Q}^{(2)},\dots,\Gamma_{Q}^{(K)}\right)$,
where each $\Gamma_Q^{(k)}$ represents a trajectory in the $k$-th replica.

\subsection{Quantum replica trade-off relations}

As in the classical case, it is straightforward to derive trade-off relations in the quantum replica process. 
Specifically, it is known that the following relation holds \cite{Hasegawa:2023:BulkBoundaryBoundNC}:
\begin{align}
    \frac{\mathrm{Var}[N_{Q}]}{\tau^{2}\left(\partial_{\tau}\mathbb{E}[N_{Q}]\right)^{2}}\ge\frac{1}{\mathcal{B}(\tau)}.
    \label{eq:QKUR_derivative}
\end{align}
Moreover, when the system is in a steady state, Eq.~\eqref{eq:QKUR_derivative} reduces to \cite{Hasegawa:2020:QTURPRL}
\begin{align}
    \frac{\mathrm{Var}[N_{Q\sharp}]}{\mathbb{E}[N_{Q\sharp}]^{2}}\ge\frac{1}{\mathcal{B}(\tau)}
    \label{eq:QKUR_steadystate}
\end{align}
which holds for the counting observable $N_{Q\sharp}(\Gamma_Q)$ defined in Eq.~\eqref{eq:quantum_counting_def}. 
In Eqs.~\eqref{eq:QKUR_derivative} and \eqref{eq:QKUR_steadystate}, $\mathcal{B}(\tau)$ denotes the quantum dynamical activity within the interval $[0,\tau]$ \cite{Hasegawa:2020:QTURPRL,Hasegawa:2023:BulkBoundaryBoundNC,Nishiyama:2024:ExactQDAPRE}. 
The classical dynamical activity [Eq.~\eqref{eq:dynamical_activity_def}] quantifies the extent of activity in the dynamics. 
For classical Markov processes, the state changes occur only when there are jumps. 
However, in the quantum scenario, state changes can occur even without jumps due to the coherent contribution arising from the $-i[H,\rho]$ term in Eq.~\eqref{eq:Lindblad_op}. 
The quantum dynamical activity plays a central role in trade-off relations in quantum thermodynamics
\cite{Hasegawa:2020:QTURPRL,Hasegawa:2023:BulkBoundaryBoundNC,
Nakajima:2023:SLD,Nishiyama:2024:ExactQDAPRE,Nishiyama:2024:OpenQuantumRURJPA}.
For the explicit expression of $\mathcal{B}(\tau)$, see Appendix~\ref{sec:QDA}. 

In the same way as in Eq.~\eqref{eq:transient_KUR}, from Eq.~\eqref{eq:QKUR_derivative}, we derive the following bound for the $K$-replica system:
\begin{align}
    \frac{\mathrm{Var}[N_{Q}^{(\otimes K)}]}{\tau^{2}\left(\partial_{\tau}\mathbb{E}[N_{Q}^{(\otimes K)}]\right)^{2}}\ge\frac{1}{K\mathcal{B}(\tau)}.
    \label{eq:transient_quantum_KUR}
\end{align}
Similarly to the classical counterpart [Eq.~\eqref{eq:transient_KUR}], the right-hand side of Eq.~\eqref{eq:transient_quantum_KUR} depends on the 
quantum dynamical activity $\mathcal{B}(\tau)$. 
The left-hand side of Eq.~\eqref{eq:transient_quantum_KUR} comprises the observable $N_{Q}^{(\otimes K)}(\Gamma_Q^{(\otimes K)}$, which is an arbitrary function of a quantum trajectory $\Gamma_Q^{(\otimes K)}$.
Moreover, for the observable $N_{Q\circ}^{(\otimes K)}$, which requires the condition given in Eq.~\eqref{eq:Nqc_condition}, the following relation holds using the result of Ref.~\cite{Hasegawa:2024:ConcentrationIneqPRL} (see Appendix~\ref{sec:quantum_replica_tradeoff_derivation} for details):
\begin{align}
    \frac{\mathbb{E}\left[|N_{Q\circ}^{(\otimes K)}|^{s}\right]^{r/(s-r)}}{\mathbb{E}\left[|N_{Q\circ}^{(\otimes K)}|^{r}\right]^{s/(s-r)}}\geq\left(1-\cos\left[\frac{1}{2}\int_{0}^{\tau}\frac{\sqrt{\mathcal{B}(t)}}{t}dt\right]^{2K}\right)^{-1},
    \label{eq:main_quantum_ENK}
\end{align}
where 
\begin{align}
    0\le\frac{1}{2}\int_{0}^{\tau}\frac{\sqrt{\mathcal{B}(t)}}{t}dt\le\frac{\pi}{2}.
    \label{eq:AQ_range_condition}
\end{align}
For the classical case, the right-hand side depends solely on the initial dynamical activity $\mathfrak{a}(t=0)$. 
On the other hand, $\mathcal{B}(\tau)$ depends on the dynamics within the interval $[0,\tau]$.

\subsection{Entropic bound on quantum trajectory observable}

We aim to quantify the uncertainty associated with quantum trajectory observables. To achieve this, we analyze the probability distribution of $N_{Q}(\Gamma_Q)$. We derive an upper bound for the Tsallis entropy of this distribution, denoted by $P(N_{Q})$.
Following the classical derivation detailed in Appendix~\ref{sec:derivation_of_timederivative}, we obtain
\begin{align}
    \left|\frac{\partial}{\partial\tau}H_{q}^{\mathrm{T}}\left[P\left(N_{Q}\right)\right]\right|\leq\frac{\sqrt{q\mathcal{B}(\tau)}}{2\tau(q-1)},
    \label{eq:dt_HT_traj_upperbound_quantum}
\end{align}
where $q=2,3,\cdots$. 
Equation~\eqref{eq:dt_HT_traj_upperbound_quantum} holds for an arbitrary observable $N_{Q}(\Gamma_Q)$ of the quantum trajectory $\Gamma_{Q}$. 
Again, for $N_{Q\circ}$, which additionally satisfies the condition of Eq.~\eqref{eq:Nqc_condition}, the following relation holds:
\begin{align}
    H_{q}^{\mathrm{T}}[P(N_{Q\circ})]\le\frac{1}{q-1}\sin\left(\frac{1}{2}\int_{0}^{\tau}\frac{\sqrt{q\mathcal{B}(t)}}{t}dt\right)^{2},
    \label{eq:HT_traj_upperbound_quantum}
\end{align}
where $q=2,3,\cdots$, and the following condition should be satisfied:
\begin{align}
    0\le\frac{1}{2}\int_{0}^{\tau}\frac{\sqrt{q\mathcal{B}(t)}}{t}dt\le\frac{\pi}{2}.
    \label{eq:AQ_range_condition2}
\end{align}
Equations~\eqref{eq:dt_HT_traj_upperbound_quantum} and \eqref{eq:HT_traj_upperbound_quantum} are quantum generalizations of Eqs.~\eqref{eq:dt_HT_traj_upperbound} and \eqref{eq:HT_traj_upperbound}, respectively, and thus their derivations are the same as those of Eqs.~\eqref{eq:dt_HT_traj_upperbound} and \eqref{eq:HT_traj_upperbound}. 
From Eq.~\eqref{eq:main_quantum_ENK}, we obtain the following relation (see Appendix~\ref{sec:quantum_entropic_bound_derivation} for details):
\begin{align}
H_{q}^{\mathrm{T}}[P(N_{Q\circ})]\le\frac{1-\cos\left[\frac{1}{2}\int_{0}^{\tau}\frac{\sqrt{\mathcal{B}(t)}}{t}dt\right]^{2q}}{q-1},
    \label{eq:quantum_Renyi_bound}
\end{align}
where $q = 2,3,\cdots$.
Equation~\eqref{eq:quantum_Renyi_bound} is a quantum generalization of Eq.~\eqref{eq:tsallis_alpha_bound_traj}. 
Equations~\eqref{eq:HT_traj_upperbound_quantum} and \eqref{eq:quantum_Renyi_bound} are similar in that the left-hand side is identical and the right-hand sides comprise the quantum dynamical activity.
It can be shown that Eq.~\eqref{eq:quantum_Renyi_bound} provides a tighter bound. 
In the classical case, 
the left-hand side of Eq.~\eqref{eq:tsallis_alpha_bound_traj} is the Tsallis entropy of the observable $N_\circ$, and the right-hand side depends on the dynamical activity $\mathfrak{a}(t=0)$ at time $t=0$.
On the other hand, in the quantum case, the right-hand side of Eq.~\eqref{eq:quantum_Renyi_bound} depends on the time integral of the quantum dynamical activity $\mathcal{B}(t)$.

\begin{figure*}
\includegraphics[width=0.9\linewidth]{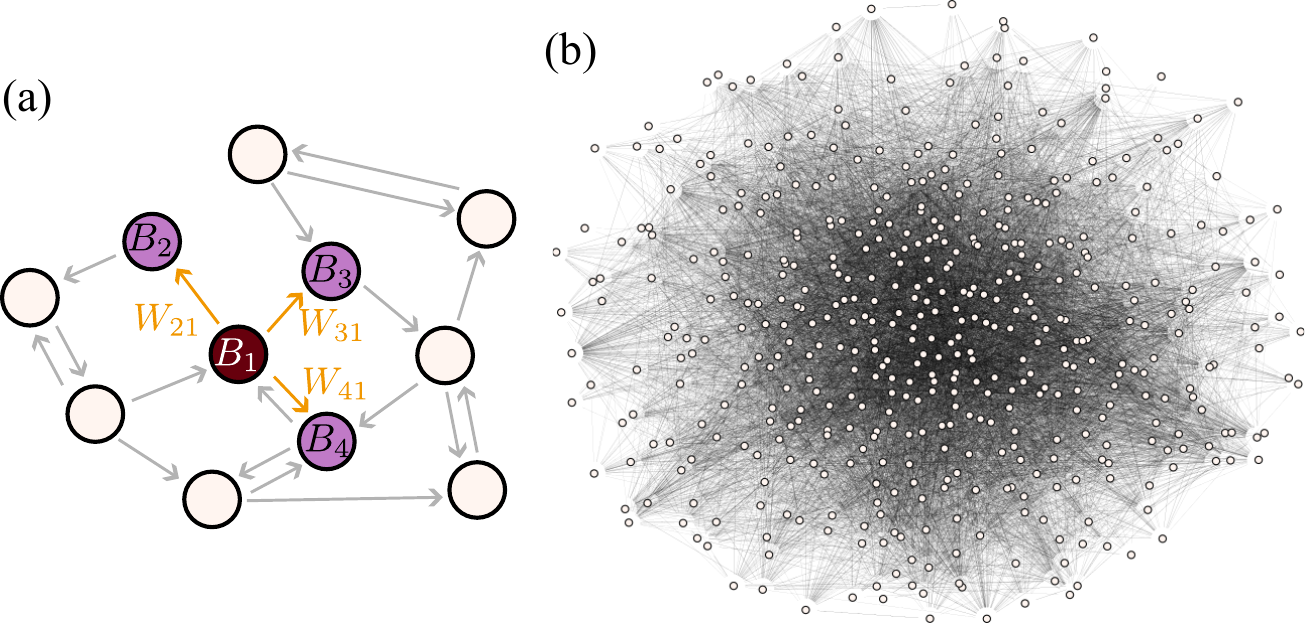}
\caption{
Description of networks. 
(a) Illustration of the dependence of the entropic bounds for the state distribution [Eqs.~\eqref{eq:renyi_alpha_bound_state} and \eqref{eq:tsallis_alpha_bound_state}].  
The entropic bounds shown in Eqs.~\eqref{eq:renyi_alpha_bound_state} and \eqref{eq:tsallis_alpha_bound_state} depend on $\sum_{\mu(\mu\ne\mu_{0})}W_{\mu\mu_{0}}$, where $B_{\mu_0}$ is the initial state, i.e., the local escape rate from $B_{\mu_0}$.
The local escape rate from $B_{\mu_0}$ is the sum of transition rates to the nodes reachable from the initial state in one jump.
Suppose the initial state is $B_1$, which is colored dark red. Then, the reachable states are represented by purple nodes. Therefore, the local escape rate becomes
$\sum_{\mu(\mu\ne\mu_{0})}W_{\mu\mu_{0}} = W_{21}+W_{31}+W_{41}$. 
(b) Twitter interaction network used in our numerical simulations. The data \cite{Fink:2023:ViralCentrality,Fink:2023:CongressTwitterInfluence}, available from \cite{Sprint:2022:CongressionalTwitterNetwork,Leskovec:2014:SNAP}, represent interactions among members of the 117th United States Congress as a directed, weighted graph. Edge weights are empirically estimated transmission probabilities. The network contains $475$ nodes and $13{,}289$ edges.
}
\label{fig:local_escape_rate}
\end{figure*}

\section{Numerical simulation}

We run numerical simulations to verify the derived bounds. 
We use Twitter interaction network data \cite{Fink:2023:ViralCentrality,Fink:2023:CongressTwitterInfluence},
which are available from \cite{Sprint:2022:CongressionalTwitterNetwork,Leskovec:2014:SNAP}. 
This dataset captures interactions among members of the 117th United States Congress as a directed, weighted social network. The overall network configuration is shown in Fig.~\ref{fig:local_escape_rate}. Each edge weight represents an empirically estimated transmission probability from one member to another. This probability was computed by counting how often a member's tweets are retweeted, quote-tweeted, replied to, or mentioned by another member and then normalizing that total by the number of tweets produced by the potential influencer. This network is well suited for studying information diffusion and influence dynamics in social systems.
In total, the number of nodes is $475$, and the number of edges is $13,289$.
We construct a continuous-time Markov process from the dataset, assigning transition rates based on the transmission probabilities. Because these probabilities are very small, we approximate the transition rates by using the corresponding transition probabilities.
We analyze the degree of information diffusion by computing the generalized entropies $H_q^\mathrm{T}$ or $H_\alpha^\mathrm{R}$ and their upper bounds.

\begin{figure}
\includegraphics[width=0.9\linewidth]{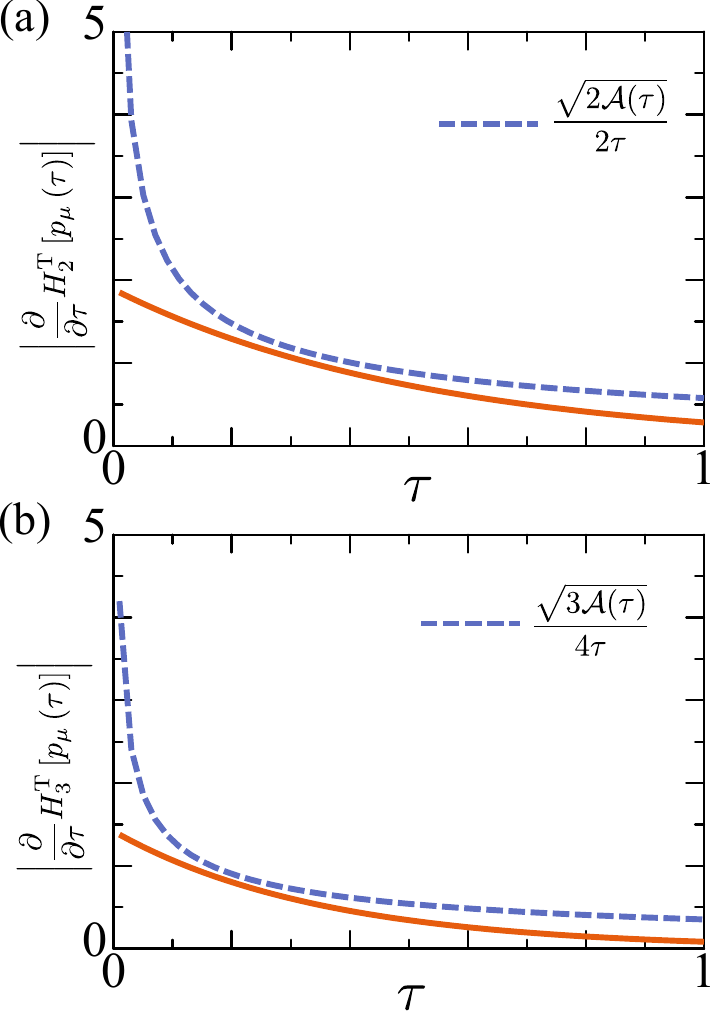}
\caption{
Numerical verification of the time derivative of the Tsallis entropy bound for the state distribution shown in Eq.~\eqref{eq:dt_HT_state_upperbound} using the Twitter interaction network. 
The time derivative of the Tsallis entropy, $\left|\partial_{\tau}H_{q}^{\mathrm{T}}\left[p_{\mu}\left(\tau\right)\right]\right|$, is plotted as a function of $\tau$ for (a) $q=2$ and (b) $q=3$ with solid lines. 
The dashed lines denote the upper bound in Eq.~\eqref{eq:dt_HT_state_upperbound}, which is (a) $\sqrt{2\mathcal{A}(\tau)}/(2\tau)$ and (b) $\sqrt{3\mathcal{A}(\tau)}/(4\tau)$. 
The initial state is $p_{\mu_0}(0)=1$, where $B_{\mu_0}$ is the node with the maximum out-degree in the Twitter interaction network. 
}
\label{fig:tsallis_statediff_time}
\end{figure}

We begin by examining the bound in Eq.~\eqref{eq:dt_HT_state_upperbound}. To quantify how information spreads through the network, we use the Tsallis entropy $H_q^\mathrm{T}[p_\mu(\tau)]$. Our focus is on the diffusion speed, measured by $\left|\partial_{\tau}H_{q}^{\mathrm{T}}\left[p_{\mu}\left(\tau\right)\right]\right|$. For this calculation, we assume that the initial state is the node with the maximum out-degree in the network.
In Fig.~\ref{fig:tsallis_statediff_time}, we plot $\left|\partial_{\tau}H_{q}^{\mathrm{T}}\left[p_{\mu}\left(\tau\right)\right]\right|$ as a function of $\tau$ for (a) $q=2$ and (b) $q=3$, where the dashed line denotes the upper bound $\sqrt{q\mathcal{A}(\tau)}/(2\tau(q-1))$ given on the right-hand side of Eq.~\eqref{eq:dt_HT_state_upperbound}. 
From Fig.~\ref{fig:tsallis_statediff_time}, it can be seen that, for both $q=2$ and $q=3$, the derivative values of the Tsallis entropy fall below the upper bound, confirming that Eq.~\eqref{eq:dt_HT_state_upperbound} holds.
The derivative value of the Tsallis entropy is larger in the region where $\tau$ is small. This indicates that diffusion occurs more rapidly in the early stages, when $\tau$ is small.
Furthermore, in both cases, a gap from the upper bound is observed in the region where $\tau$ is small.

\begin{figure}
\includegraphics[width=0.9\linewidth]{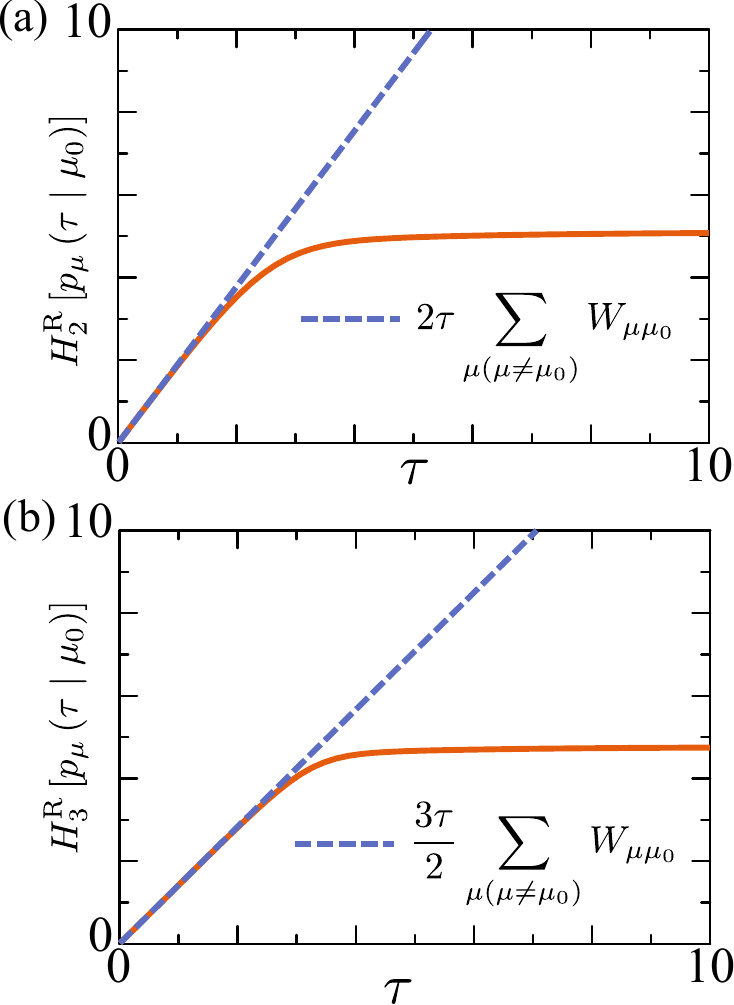}
\caption{
Numerical verification of the R\'enyi entropy bound for the state distribution shown in Eq.~\eqref{eq:renyi_alpha_bound_state} using the Twitter interaction network. 
The R\'enyi entropy $H_{\alpha}^{\mathrm{R}}[p_{\mu}(\tau\mid\mu_{0})]$ is plotted as a function of $\tau$ for (a) $\alpha=2$ and (b) $\alpha=3$ using solid lines. 
The dashed lines denote the upper bounds in Eq.~\eqref{eq:renyi_alpha_bound_state}, which are (a) $2\tau\sum_{\mu(\mu\ne\mu_{0})}W_{\mu\mu_{0}}$ and (b) $(3\tau/2)\sum_{\mu(\mu\ne\mu_{0})}W_{\mu\mu_{0}}$. 
The initial state is $p_{\mu_0}(0)=1$, where $B_{\mu_0}$ is the node with the maximum out-degree in the Twitter interaction network. 
}
\label{fig:renyi_state_time}
\end{figure}

Next, we examine the entropy at time $\tau$ and its upper bound.
Here, we use the R\'enyi entropy [Eq.~\eqref{eq:Renyi_ent_def}] rather than the Tsallis entropy to quantify the degree of diffusion. 
The reason for employing the R\'enyi entropy is that, in the upper bound of the Tsallis entropy [Eq.~\eqref{eq:tsallis_alpha_bound_state}], the dynamical activity and $\tau$ are included in the exponent,
which causes the upper bound to rapidly converge to $1/(q-1)$.
Again, we assume that the initial state is the node with the maximum out-degree. 
In Fig.~\ref{fig:renyi_state_time}, we plot $H_{\alpha}^{\mathrm{R}}[p_{\mu}(\tau\mid\mu_{0})]$ as a function of $\tau$ for (a) $\alpha=2$ and (b) $\alpha=3$, where the dashed line denotes the upper bound $\alpha\tau/(\alpha-1)\sum_{\mu(\mu\ne\mu_{0})}W_{\mu\mu_{0}}$ given on the right-hand side of Eq.~\eqref{eq:renyi_alpha_bound_state}. 
It can be seen that the bound represented by Eq.~\eqref{eq:renyi_alpha_bound_state} provides a very tight upper bound up to approximately $\tau=2$. In the region where $\tau$ is greater than $3$, while the entropy value is nearly saturated, the upper bound increases linearly with respect to $\tau$, resulting in a gap between the upper bound and the entropy value. The behaviors for $\alpha=2$ and $\alpha=3$ are similar, but around the region where the slope of the entropy changes, $\alpha=3$ changes more steeply.

\begin{figure}
\includegraphics[width=0.9\linewidth]{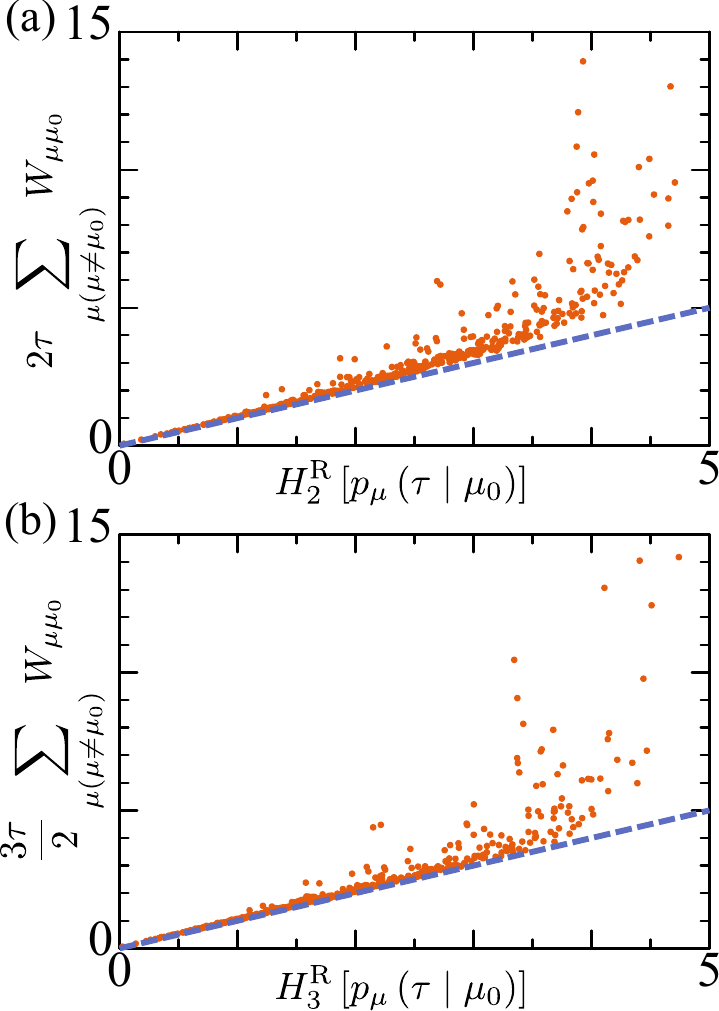}
\caption{
Numerical verification of the R\'enyi entropy bound for the state distribution shown in Eq.~\eqref{eq:renyi_alpha_bound_state} using the Twitter interaction network. 
The R\'enyi entropy $H_{\alpha}^{\mathrm{R}}[p_{\mu}(\tau\mid\mu_{0})]$
versus $\alpha\tau/(\alpha-1)\sum_{\mu(\mu\ne\mu_{0})}W_{\mu\mu_{0}}$ at time $\tau = 10$
is plotted for (a) $\alpha=2$ and (b) $\alpha=3$. 
Each point corresponds to a different initial state, and we use every $475$ node as the initial state. 
The dashed lines denote the equality case in Eq.~\eqref{eq:renyi_alpha_bound_state}. 
}
\label{fig:renyi_state_vs}
\end{figure}

Finally, we compare the left and right sides of Eq.~\eqref{eq:renyi_alpha_bound_state} for different initial settings. 
We calculate $H_{\alpha}^{\mathrm{R}}[p_{\mu}(\tau)]$ and $\alpha\tau/(\alpha-1)\sum_{\mu(\mu\ne\mu_{0})}W_{\mu\mu_{0}}$, which are the left and right sides of Eq.~\eqref{eq:renyi_alpha_bound_state} at time $\tau = 10$. 
There are $475$ nodes in the network, and using all of these nodes as initial states, the results for each of the $475$ configurations are plotted as points on Fig.~\ref{fig:renyi_state_vs} for (a) $\alpha=2$ and (b) $\alpha=3$. 
The equality case of Eq.~\eqref{eq:renyi_alpha_bound_state} is shown by the dashed lines in Fig.~\ref{fig:renyi_state_vs}. 
Because all the points in Figs.~\ref{fig:renyi_state_vs}(a) and (b) are above the dashed lines, we can numerically validate Eq.~\eqref{eq:renyi_alpha_bound_state}.

\section{Conclusion}

We have introduced \emph{replica Markov processes}, collections of $K$ independent, identical copies of a Markov process, as a systematic framework for deriving thermodynamic trade-off relations that constrain nonlinear functionals of probability distributions.
By applying existing activity-based trade-off relations [Eqs.~\eqref{eq:transient_KUR} and \eqref{eq:main_result1}] to suitably chosen observables in the replica process, nonlinear quantities of the original single process are expressed as linear expectations in the enlarged state space. The replicas serve as purely virtual constructs, analogous to multi-copy measurements in quantum information and the replica trick in spin-glass theory, so that the final results are statements about the single process alone.
Focusing on entropic measures of uncertainty, we have obtained several families of bounds:
\begin{enumerate}
\item \textit{Trajectory-observable entropies.}
Upper bounds on the time derivative [Eq.~\eqref{eq:dt_HT_traj_upperbound}] and the value [Eq.~\eqref{eq:HT_traj_upperbound}] of the Tsallis entropy for the distribution of a general trajectory observable, expressed in terms of the time-integrated dynamical activity $\mathcal{A}(\tau)$.
Additionally, R\'enyi and Tsallis entropy bounds [Eqs.~\eqref{eq:renyi_alpha_bound} and \eqref{eq:tsallis_alpha_bound_traj}] controlled by the initial dynamical activity $\mathfrak{a}(t{=}0)$ have been derived.

\item \textit{State-distribution entropies and network diffusion.}
Analogous bounds on the time derivative [Eq.~\eqref{eq:dt_HT_state_upperbound}] and the value [Eq.~\eqref{eq:HT_state_upperbound}] of the Tsallis entropy for the state probability distribution $p_\mu(\tau)$, which quantifies the degree of diffusion on a network.
Notably, the R\'enyi and Tsallis entropy bounds [Eqs.~\eqref{eq:renyi_alpha_bound_state} and \eqref{eq:tsallis_alpha_bound_state}] depend only on the observation time $\tau$ and the local escape rate from the initial node, $\sum_{\mu(\mu\ne\mu_0)}W_{\mu\mu_0}$, irrespective of the global network size or topology.

\item \textit{Extreme-value observables.}
We have demonstrated that the replica construction naturally extends to other nonlinear statistics. In particular, trade-off relations for the maximum of $M$ independent realizations [Eqs.~\eqref{eq:transient_KUR_Mmax}--\eqref{eq:NM_max_Tsallis_upperbound2}] have been obtained, showing that the entropic uncertainty of extreme-value observables is similarly governed by the dynamical activity.

\item \textit{Quantum extension.}
The entire framework has been generalized to continuously monitored open quantum systems described by Lindblad dynamics. The resulting bounds [Eqs.~\eqref{eq:dt_HT_traj_upperbound_quantum}, \eqref{eq:HT_traj_upperbound_quantum}, and \eqref{eq:quantum_Renyi_bound}] take the same structural form as their classical counterparts, with the classical dynamical activity replaced by the quantum dynamical activity $\mathcal{B}(\tau)$, which additionally accounts for coherent contributions to the dynamics.
\end{enumerate}

Numerical simulations on a real-world Twitter interaction network among members of the 117th United States Congress have confirmed the validity and tightness of the derived bounds, particularly the R\'enyi entropy bound for the state distribution [Eq.~\eqref{eq:renyi_alpha_bound_state}].
Our framework extends the scope of activity-based trade-off relations from linear functionals of probability distributions to genuinely nonlinear information-theoretic quantities, which are inaccessible to conventional methods.

\appendix

\section{Derivation of Eqs.~\eqref{eq:dt_HT_traj_upperbound}--\eqref{eq:HT_traj_upperbound}\label{sec:derivation_of_timederivative}}

Let us consider the observable:
\begin{align}
    &\Lambda^{(\otimes K)}(\Gamma^{(\otimes K)})\nonumber\\&\equiv\begin{cases}
1 & N^{(1)}(\Gamma^{(1)})=N^{(2)}(\Gamma^{(2)})=\cdots=N^{(K)}(\Gamma^{(K)})\\
0 & \mathrm{otherwise}
\end{cases}.
\label{eq:Nx_match_def}
\end{align}
$\Lambda^{(\otimes K)}$ returns 1 only when the values of the observable $N$ in the $K$ processes agree, $N^{(1)}(\Gamma^{(1)})=N^{(2)}(\Gamma^{(2)})=\cdots=N^{(K)}(\Gamma^{(K)})$.
Let $ \mathcal{Z}$ be the set of all possible values that the observable $ N(\Gamma) $ can take.
The expectation and variance of Eq.~\eqref{eq:Nx_match_def} are
\begin{align}
    \mathbb{E}[\Lambda^{(\otimes K)}(\Gamma^{(\otimes K)})]&=\sum_{z\in\mathcal{Z}}P(z)^{K},\label{eq:expectation_Lambda}\\
    \mathrm{Var}[\Lambda^{(\otimes K)}(\Gamma^{(\otimes K)})]&=\sum_{z\in\mathcal{Z}}P(z)^{K}\left(1-\sum_{z\in\mathcal{Z}}P(z)^{K}\right),
    \label{eq:variance_Lambda}
\end{align}
where we abbreviate $P(z) \equiv P(N(\Gamma)=z)$. 
Substituting Eqs.~\eqref{eq:expectation_Lambda} and \eqref{eq:variance_Lambda} into Eq.~\eqref{eq:transient_KUR}, we obtain
\begin{align}
    \frac{\left|\partial_{\tau}\sum_{z}P(z)^{K}\right|}{\sqrt{\sum_{z}P(z)^{K}\left(1-\sum_{z}P(z)^{K}\right)}}\le\frac{\sqrt{K\mathcal{A}(\tau)}}{\tau}.
    \label{eq:dtau_pmuK_KAtau}
\end{align}
For an arbitrary function $f(\tau)$, the following relation holds:
\begin{align}
    \frac{d}{d\tau}\arcsin\left(2f-1\right)=\frac{1}{\sqrt{f(1-f)}}\left(\frac{df}{d\tau}\right).
    \label{eq:arcsin_ftau}
\end{align}
Using Eq.~\eqref{eq:arcsin_ftau} in Eq.~\eqref{eq:dtau_pmuK_KAtau}, we have
\begin{align}
    \left|\frac{d}{d\tau}\arcsin\left(2\sum_{z\in\mathcal{Z}}P(z)^{K}-1\right)\right|\leq\frac{\sqrt{K\mathcal{A}(\tau)}}{\tau}.
    \label{eq:dtau_arcsin_KAtau}
\end{align}
Recall that the following relation holds for $|f(\tau)|<1$:
\begin{align}
    \left|\frac{d}{d \tau} f(\tau)\right| \leq \frac{1}{2}\left|\frac{d}{d \tau} \arcsin (2 f(\tau)-1)\right|.
    \label{eq:time_derivative_ineq}
\end{align}
Using Eq.~\eqref{eq:time_derivative_ineq} with Eq.~\eqref{eq:dtau_arcsin_KAtau} yields
\begin{align}
    \left|\frac{d}{d\tau}\sum_{z\in\mathcal{Z}}P(z)^{K}\right|\le\frac{\sqrt{K\mathcal{A}(\tau)}}{2\tau},
    \label{eq:dtau_pmuK_KAtau_2tau}
\end{align}
which proves Eq.~\eqref{eq:dt_HT_traj_upperbound} in the main text. 

Next, we derive Eq.~\eqref{eq:HT_traj_upperbound}.
Instead of $N(\Gamma)$, we consider $N_\circ(\Gamma)$ in Eq.~\eqref{eq:dtau_arcsin_KAtau}.
While $N(\Gamma)$ is an arbitrary function of $\Gamma$, $N_\circ(\Gamma)$ has the additional assumption given in Eq.~\eqref{eq:zero_condition}. 
Let $ \mathcal{Z}_\circ$ be the set of all possible values that the observable $ N_\circ(\Gamma) $ can take.
By integrating 
Eq.~\eqref{eq:dtau_arcsin_KAtau} from $t=0$ to $t=\tau$ and using the triangle inequality, we obtain
\begin{align}
&\left|\arcsin\left(2\sum_{z\in\mathcal{Z}_\circ}P(N_{\circ}=z)^{K}-1\right)-\frac{\pi}{2}\right|\nonumber\\&\leq\int_{0}^{\tau}\frac{\sqrt{K\mathcal{A}(t)}}{t}dt.
    \label{eq:arcsin_pmu}
\end{align}
In Eq.~\eqref{eq:arcsin_pmu}, we used the fact that $N_\circ = 0$ at $t=0$ due to the condition of Eq.~\eqref{eq:zero_condition} (recall that, at time $t=0$, there is no jump).
Therefore, the probability is concentrated on $P(N_\circ=0)=1$ at time $t=0$. 
When $0\le(1/2)\int_{0}^{\tau}\sqrt{K\mathcal{A}(t)}/t\,dt\le\pi/2$, we obtain from Eq.~\eqref{eq:arcsin_pmu}
\begin{align}
    \sum_{z\in\mathcal{Z}_{\circ}}P(N_{\circ}=z)^{K}\geq\cos\left(\frac{1}{2}\int_{0}^{\tau}\frac{\sqrt{K\mathcal{A}(t)}}{t}dt\right)^{2},
    \label{eq:pmuK_cos2}
\end{align}
which yields Eq.~\eqref{eq:HT_traj_upperbound} in the main text. 

\section{Derivation of Eqs.~\eqref{eq:dt_HT_state_upperbound}, \eqref{eq:dt_HT_state_upperbound_integration}, and \eqref{eq:HT_state_upperbound}\label{sec:derivation_of_timederivative_state}}

By adopting the derivation above, we can obtain the bounds shown in Eqs.~\eqref{eq:dt_HT_state_upperbound}, \eqref{eq:dt_HT_state_upperbound_integration}, and \eqref{eq:HT_state_upperbound}. 
Since $N(\Gamma)$ can be an arbitrary function of $\Gamma$, we can take
\begin{align}
    N(\Gamma) = X(\tau),
    \label{eq:N_equal_Xtau}
\end{align}
where $X(\tau)$ is the state of the Markov process at time $t=\tau$.
Using Eq.~\eqref{eq:N_equal_Xtau} in Eq.~\eqref{eq:dtau_pmuK_KAtau_2tau}, we obtain
\begin{align}
    \left|\frac{d}{d\tau}\sum_{\mu}p_{\mu}(\tau)^{K}\right|\le\frac{\sqrt{K\mathcal{A}(\tau)}}{2\tau},
    \label{eq:dtau_pmuK_ineq_KAtau}
\end{align}
Moreover, integrating Eq.~\eqref{eq:dtau_pmuK_ineq_KAtau} from $t=0$ to $t=\tau$, we obtain
\begin{align}
    \left|\sum_{\mu}p_{\mu}(\tau)^{K}-\sum_{\mu}p_{\mu}(0)^{K}\right|\leq\int_{0}^{\tau}\frac{\sqrt{K\mathcal{A}(t)}}{2t}dt,
    \label{eq:dtau_pmuK_ineq_KAtau_integration}
\end{align}
which derives Eq.~\eqref{eq:dt_HT_state_upperbound_integration} in the main text. 
When the initial state of the process is fixed to $B_{\mu_0}$, implying $p_{\mu_0}(0)=1$, we can derive a tighter inequality than Eq.~\eqref{eq:dtau_pmuK_ineq_KAtau_integration}. 
Specifically, substituting Eq.~\eqref{eq:N_equal_Xtau} into Eq.~\eqref{eq:dtau_arcsin_KAtau}, we obtain
\begin{align}
    \left|\arcsin\left(2\sum_{\mu}p_{\mu}(\tau\mid\mu_{0})^{K}-1\right)-\frac{\pi}{2}\right|\leq\int_{0}^{\tau}\frac{\sqrt{K\mathcal{A}(t)}}{t}dt,
    \label{eq:arcsin_pmuK_ineq_KAtau_integration}
\end{align}
which yields
\begin{align}
    \sum_{\mu}p_{\mu}(\tau\mid\mu_{0})^{K}\geq\cos\left(\frac{1}{2}\int_{0}^{\tau}\frac{\sqrt{K\mathcal{A}(t)}}{t}dt\right)^{2},
    \label{eq:pmuK_cos_KAtau_sq}
\end{align}
where $0\le(1/2)\int_{0}^{\tau}\sqrt{K\mathcal{A}(t)}/t\,dt\le\pi/2$ should be satisfied. 
Equation~\eqref{eq:pmuK_cos_KAtau_sq} derives Eq.~\eqref{eq:HT_state_upperbound} in the main text.

\section{Derivation of Eq.~\eqref{eq:main_result1}\label{sec:classical_nojump_KUR}}

The derivation of the main result [Eq.~\eqref{eq:main_result1}] is based on Ref.~\cite{Hasegawa:2024:ConcentrationIneqPRL} and 
we extend the bound to the case including $K$ replicas.

Suppose that the initial state in the trajectory $\Gamma$ is $B_{\mu_0}$. 
The probability of the trajectory $\Gamma$ [Eq.~\eqref{eq:trajectory_single_def}] is given by \cite{Seifert:2012:FTReview}
\begin{align}
    \mathcal{P}\left(\Gamma\right)=p_{\mu_0}(t=0)\mathcal{P}\left(\Gamma\mid\mu_{0}\right),
    \label{eq:path_prob_def}
\end{align}
where $\mathcal{P}(\Gamma \mid\mu_0)$ is the trajectory probability given that the initial state is $B_{\mu_0}$ and $X_j = B_{\mu_j}$:
\begin{align}
    \mathcal{P}\left(\Gamma\mid\mu_{0}\right)=\prod_{j=1}^{J}W_{\mu_{j}\mu_{j-1}}e^{-\sum_{j=0}^{J}\left(t_{j+1}-t_{j}\right)R\left(\mu_{j}\right)},
    \label{eq:P_Gamma_mu0_theta}
\end{align}
where $t_{J+1}\equiv \tau$. 
Here, $R(\mu)$ is defined by
\begin{align}
    R(\mu)\equiv\sum_{\mu^{\prime}(\neq\mu)}W_{\mu^{\prime}\mu}.
    \label{eq:Markov_R_def}
\end{align}
Let $\mathfrak{p}(\tau)$ be the probability that there is no jump within the interval $[0,\tau]$.
From Eq.~\eqref{eq:path_prob_def},
$\mathfrak{p}(\tau)$ is expressed as
\begin{align}
\mathfrak{p}(\tau)&=\sum_{\mu_{0}}p_{\mu_{0}}(t=0)e^{-\tau R\left(\mu_{0}\right)}.
\label{eq:ptau_expression}
\end{align}
Using the Jensen inequality, we obtain 
\begin{align}
    \mathfrak{p}(\tau)&\ge e^{-\tau\sum_{\mu_{0}}p_{\mu_{0}}(t=0)R\left(\mu_{0}\right)}\nonumber\\&=e^{-\tau\mathfrak{a}(t=0)}.
    \label{eq:nojump_jensen}
\end{align}

Next, we consider the trajectory $\Gamma^{(\otimes K)}$ in the $K$ replica processes.
Let $\mathfrak{p}^{(\otimes K)}(\tau)$ be the probability that there is no jump within the interval $[0,\tau]$ in the $K$-replica case.
Since the replicas are independent processes, from Eq.~\eqref{eq:nojump_jensen}, $\mathfrak{p}^{(\otimes K)}(\tau)$ is bounded from below by
\begin{align}
    \mathfrak{p}^{(\otimes K)}(\tau)&\ge\left[e^{-\tau\sum_{\mu_{0}}p_{\mu_{0}}(t=0)R\left(\mu_{0}\right)}\right]^{K}\nonumber\\&=e^{-K\tau\mathfrak{a}(t=0)}.
    \label{eq:nojump_K_lowerbound}
\end{align}
We can relate $\mathfrak{p}^{(\otimes K)}(\tau)$ to the probability $P\left[N_{\circ}^{(\otimes K)}(\Gamma^{(\otimes K)})=0\right]$, which is the probability that the observable $N_{\circ}^{(\otimes K)}$ vanishes.  
If no jump occurs in the interval $[0,\tau]$, then the observable satisfies $N_{\circ}^{(\otimes K)}(\Gamma^{(\otimes K)})=0$ by the condition in Eq.~\eqref{eq:NK_null_condition}.
The converse, however, does not necessarily hold. Therefore,
the probability that $N_{\circ}^{(\otimes K)}(\Gamma^{(\otimes K)})=0$ is bounded from below by
\begin{align}
    P\left[N_{\circ}^{(\otimes K)}(\Gamma^{(\otimes K)})=0\right]\geq\mathfrak{p}^{(\otimes K)}(\tau).
    \label{eq:PNK_ineq_ptau}
\end{align}
Let $X$ be a random variable.
According to Ref.~\cite{Valentin:2007:TailProb},
the following relation holds:
\begin{align}
    P(|X|>b) \geq \frac{\left(\mathbb{E}\left[|X|^r\right]-b^r\right)^{s /(s-r)}}{\mathbb{E}\left[|X|^s\right]^{r /(s-r)}},
    \label{eq:Petrov_ineq}
\end{align}
where $0<r<s$, $b \geq 0$, and $b^r \leq \mathbb{E}\left[|X|^r\right]$ should be satisfied. 
Using Eq.~\eqref{eq:Petrov_ineq} with $b=0$ for the random variable $N_{\circ}^{(\otimes K)}$, we obtain
\begin{align}
    \frac{\mathbb{E}\left[|N_{\circ}^{(\otimes K)}|^{r}\right]^{s/(s-r)}}{\mathbb{E}\left[|N_{\circ}^{(\otimes K)}|^{s}\right]^{r/(s-r)}}&\le P(|N_{\circ}^{(\otimes K)}|>0)\nonumber\\&=1-P(|N_{\circ}^{(\otimes K)}|=0)\nonumber\\&\le1-\mathfrak{p}^{(\otimes K)}(\tau)\nonumber\\&\le1-e^{-K\tau\mathfrak{a}(t=0)},
    \label{eq:main_result1_app}
\end{align}
where we used Eqs.~\eqref{eq:nojump_K_lowerbound} and \eqref{eq:PNK_ineq_ptau}. 
Equation~\eqref{eq:main_result1_app} is Eq.~\eqref{eq:main_result1}.

\section{Derivation of Eqs.~\eqref{eq:renyi_alpha_bound} and \eqref{eq:renyi_alpha_bound_state}\label{sec:derivation_entropic_bound}}

We show the derivation of Eqs.~\eqref{eq:renyi_alpha_bound} and \eqref{eq:renyi_alpha_bound_state}. 
We consider a particular observable to obtain inequalities for the original single process. 
Specifically, let us consider the following observable:
\begin{align}
    &\overline{\Lambda}_{\circ}^{(\otimes K)}(\Gamma^{(\otimes K)})\nonumber\\&\equiv\begin{cases}
0 & N_{\circ}^{(1)}(\Gamma^{(1)})=N_{\circ}^{(2)}(\Gamma^{(2)})=\cdots=N_{\circ}^{(K)}(\Gamma^{(K)})\\
1 & \mathrm{otherwise}
\end{cases}
\label{eq:N_match_def}
\end{align}
$\overline{\Lambda}^{(\otimes K)}_\circ$ is similar to Eq.~\eqref{eq:Nx_match_def}. However, the cases in which Eqs.~\eqref{eq:Nx_match_def} and \eqref{eq:N_match_def} return 0 and 1 are reversed.
$\overline{\Lambda}_\circ^{(\otimes K)}$ returns 0 only when the values of the observable $N_\circ$ in the $K$ processes agree, $N_{\circ}^{(1)}(\Gamma^{(1)})=N_{\circ}^{(2)}(\Gamma^{(2)})=\cdots N_{\circ}^{(K)}(\Gamma^{(K)})$.
When there is no jump in all $K$ processes, we trivially obtain
$N_{\circ}^{(1)}(\Gamma_{\varnothing}^{(1)})=N_{\circ}^{(2)}(\Gamma_{\varnothing}^{(2)})=\cdots =N_{\circ}^{(K)}(\Gamma_{\varnothing}^{(K)})=0$. 
Then the observable $\overline{\Lambda}_\circ^{(\otimes K)}$ satisfies
\begin{align}
    \overline{\Lambda}_\circ^{(\otimes K)}(\Gamma_{\varnothing}^{(\otimes K)})=0,
    \label{eq:Ntilde_null}
\end{align}
which is the condition of Eq.~\eqref{eq:NK_null_condition}. 
Then, the expectation becomes the following:
\begin{align}
    \mathbb{E}\left[\overline{\Lambda}_{\circ}^{(\otimes K)}(\Gamma^{(\otimes K)})\right]&=1-\sum_{z\in\mathcal{Z}_{\circ}}\prod_{k=1}^{K}P(N_{\circ}^{(k)}(\Gamma^{(k)})=z)\nonumber\\&=1-\sum_{z\in\mathcal{Z}_{\circ}}P(N_{\circ}(\Gamma)=z)^{K}.
    \label{eq:N_match_expectation}
\end{align}
The expectation and variance of $\overline{\Lambda}_\circ^{(\otimes K)}$ are 
\begin{align}
    \mathbb{E}[\overline{\Lambda}_{\circ}^{(\otimes K)}(\Gamma^{(\otimes K)})]&=1-\sum_{z\in\mathcal{Z}_{\circ}}P(N_{\circ}=z)^{K},\label{eq:expectation_overlineLambda}\\\mathrm{Var}[\overline{\Lambda}_{\circ}^{(\otimes K)}(\Gamma^{(\otimes K)})]&=\left(1-\sum_{z\in\mathcal{Z}_{\circ}}P(N_{\circ}=z)^{K}\right)\nonumber\\&-\left(1-\sum_{z\in\mathcal{Z}_{\circ}}P(N_{\circ}=z)^{K}\right)^{2}.
    \label{eq:variance_overlineLambda}
\end{align}
Substituting Eqs.~\eqref{eq:expectation_overlineLambda} and \eqref{eq:variance_overlineLambda} into Eq.~\eqref{eq:transient_nojump_KUR}, we obtain
\begin{align}
    \sum_{z\in\mathcal{Z}_{\circ}}P(N_{\circ}(\Gamma)=z)^{K}\ge e^{-K\tau\mathfrak{a}(t=0)}.
    \label{eq:PNc_e2taua}
\end{align}
which derives Eq.~\eqref{eq:renyi_alpha_bound} in the main text.

Next, we consider the following observable:
\begin{align}
    \overline{\Xi}^{(\otimes K)}(\Gamma^{(\otimes K)})\equiv\begin{cases}
0 & X^{(1)}(\tau)=X^{(2)}(\tau)=\cdots=X^{(K)}(\tau)\\
1 & \mathrm{otherwise}
\end{cases}.
\label{eq:Xi_obs_def}
\end{align}
Here, $X^{(k)}(\tau)$ is the state of the $k$-th process at time $t=\tau$.
Therefore, Eq.~\eqref{eq:Xi_obs_def} is $0$ if the final states of the $K$ processes agree. 
When deriving Eqs.~\eqref{eq:renyi_alpha_bound_state} and \eqref{eq:tsallis_alpha_bound_state}, we assume that the initial state of the random walk is $B_{\mu_0}$. 
This implies that when there is no jump within the interval $[0,\tau]$, the final state at time $t=\tau$ is $X^{(1)}(\tau)=X^{(2)}(\tau)=\cdots=X^{(K)}(\tau)=B_{\mu_{0}}$. 
Therefore, the observable $\overline{\Xi}^{(\otimes K)}$ satisfies
\begin{align}
    \overline{\Xi}^{(\otimes K)}(\Gamma_{\varnothing}^{(\otimes K)})=0,
    \label{eq:Xi_varnothing_zero}
\end{align}
which satisfies the condition given in Eq.~\eqref{eq:NK_null_condition}. 
The expectation of $\overline{\Xi}^{(\otimes K)}$ is given by
\begin{align}
\mathbb{E}\left[\overline{\Xi}^{(\otimes K)}(\Gamma^{(\otimes K)})\right]&=1-\sum_{\mu}\prod_{k=1}^{K}P(X^{(k)}(\tau)=B_{\mu}^{(k)})\nonumber\\&=1-\sum_{\mu}p_{\mu}(\tau\mid\mu_{0})^{K}.
    \label{eq:Xi_obs_expectation}
\end{align}
Substituting Eq.~\eqref{eq:Xi_obs_expectation} into Eq.~\eqref{eq:transient_nojump_KUR}, we obtain
\begin{align}
    \sum_{\mu}p_{\mu}(\tau\mid\mu_{0})^{K}\ge e^{-K\tau\mathfrak{a}(t=0)},
    \label{eq:Pmu_eKtaua}
\end{align}
which derives Eq.~\eqref{eq:renyi_alpha_bound_state} in the main text.

\section{Derivation of Eq.~\eqref{eq:main_quantum_ENK}\label{sec:quantum_replica_tradeoff_derivation}}

We show the derivation of Eq.~\eqref{eq:main_quantum_ENK}, which is the quantum generalization of Eq.~\eqref{eq:main_result1}. 
Similarly to the classical case, let $\mathfrak{p}_Q(\tau)$ be the probability that there is no jump within the interval $[0,\tau]$ in $\Gamma_Q$. 
From the result in Ref.~\cite{Hasegawa:2024:ConcentrationIneqPRL}, the following relation holds:
\begin{align}
    \cos\left[\frac{1}{2}\int_{0}^{\tau}\frac{\sqrt{\mathcal{B}(t)}}{t}dt\right]^{2}\leq\mathfrak{p}_{Q}(\tau),
    \label{eq:quantum_ptau_bound}
\end{align}
where the following condition should be met:
\begin{align}
    0\le\frac{1}{2}\int_{0}^{\tau}\frac{\sqrt{\mathcal{B}(t)}}{t}dt\le\frac{\pi}{2}.
    \label{eq:Btau_range}
\end{align}
Here, $\mathcal{B}(\tau)$ is the quantum dynamical activity, whose expression is expressed in Appendix~\ref{sec:QDA}. 
We consider the $K$ replica scenario, where there are $K$ identical processes. 
Let $\mathfrak{p}^{(\otimes K)}_Q(\tau)$ denote the probability that there is no jump in the $K$ replica processes.
Since each process is independent, from Eq.~\eqref{eq:quantum_ptau_bound},
$\mathfrak{p}^{(\otimes K)}_Q(\tau)$ satisfies
\begin{align}
    \cos\left[\frac{1}{2}\int_{0}^{\tau}\frac{\sqrt{\mathcal{B}(t)}}{t}dt\right]^{2K}\leq\mathfrak{p}_{Q}^{(\otimes K)}(\tau).
    \label{eq:pKtau_bound_quantum}
\end{align}
If no jump occurs in the interval $[0,\tau]$, then the observable satisfies $N_{Q\circ}^{(\otimes K)}(\Gamma_{Q\varnothing}^{(\otimes K)})=0$ by the condition in Eq.~\eqref{eq:Nqc_condition}.
The converse, however, does not necessarily hold. Therefore,
the probability that $N_{Q\circ}^{(\otimes K)}(\Gamma_{Q}^{(\otimes K)})=0$ is bounded from below by
\begin{align}
    P\left[N_{Q\circ}^{(\otimes K)}(\Gamma_{Q}^{(\otimes K)})=0\right]\geq\mathfrak{p}_{Q}^{(\otimes K)}(\tau).
    \label{eq:PNq_ineq_pqtau}
\end{align}
Using the Petrov inequality [Eq.~\eqref{eq:Petrov_ineq}], we have
\begin{align}
\frac{\mathbb{E}\left[|N_{Q\circ}^{(\otimes K)}|^{r}\right]^{s/(s-r)}}{\mathbb{E}\left[|N_{Q\circ}^{(\otimes K)}|^{s}\right]^{r/(s-r)}}&\le P(|N_{Q\circ}^{(\otimes K)}|>0)\nonumber\\&=1-P(|N_{Q\circ}^{(\otimes K)}|=0)\nonumber\\&\le1-\mathfrak{p}_{Q}^{(\otimes K)}(\tau)\nonumber\\&\le1-\cos\left[\frac{1}{2}\int_{0}^{\tau}\frac{\sqrt{\mathcal{B}(t)}}{t}dt\right]^{2K}.
    \label{eq:main_ineq_series_quantum}
\end{align}
where we used Eqs.~\eqref{eq:pKtau_bound_quantum} and \eqref{eq:PNq_ineq_pqtau}.
Equation~\eqref{eq:main_ineq_series_quantum} proves Eq.~\eqref{eq:main_quantum_ENK} in the main text.

\section{Derivation of Eq.~\eqref{eq:quantum_Renyi_bound}\label{sec:quantum_entropic_bound_derivation}}

We show the derivation
of Eq.~\eqref{eq:quantum_Renyi_bound}, which directly parallels the derivation
of Eq.~\eqref{eq:renyi_alpha_bound}. 
Specifically, let us consider the following observable for $K$ replicas:
\begin{align}
&\overline{\Lambda}_{Q\circ}^{(\otimes K)}(\Gamma_{Q}^{\otimes K})\nonumber\\&\equiv\begin{cases}
0 & N_{Q\circ}^{(1)}(\Gamma_{Q}^{(1)})=N_{Q\circ}^{(2)}(\Gamma_{Q}^{(2)})=\cdots=N_{Q\circ}^{(K)}(\Gamma_{Q}^{(K)})\\
1 & \mathrm{otherwise}
\end{cases}.
\label{eq:Nq_match_def}
\end{align}
$\overline{\Lambda}_{Q\circ}^{(\otimes K)}$ returns 0 only when the values of the observable $N_{Q\circ}$ in the $K$ processes agree, $N_{Q\circ}^{(1)}(\Gamma_{Q}^{(1)})=N_{Q\mathrm{\circ}}^{(2)}(\Gamma_{Q}^{(2)})=\cdots=N_{Q\circ}^{(K)}(\Gamma_{Q}^{(K)})$.
Then, the observable $\overline{\Lambda}_{Q\circ}^{(\otimes K)}$ satisfies
\begin{align}
    \overline{\Lambda}_{Q\circ}^{(\otimes K)}(\Gamma_{Q\varnothing}^{\otimes K})=0,
    \label{eq:Nqtilde_null}
\end{align}
which is the condition of Eq.~\eqref{eq:Nqc_condition}. 
Let $ \mathcal{Z}_{Q\circ}$ be the set of all possible values that the observable $ N_{Q\circ}(\Gamma_Q) $ can take.
The expectation becomes the following:
\begin{align}
\mathbb{E}\left[\overline{\Lambda}_{Q\circ}^{(\otimes K)}(\Gamma_{Q}^{\otimes K})\right]&=1-\sum_{z\in\mathcal{Z}_{Q\circ}}\prod_{k=1}^{K}P(N_{Q\circ}^{(k)}(\Gamma_{Q}^{(k)})=z)\nonumber\\&=1-\sum_{z\in\mathcal{Z}_{Q\circ}}P(N_{Q\circ}(\Gamma_{Q})=z)^{K}.
    \label{eq:Nq_match_expectation}
\end{align}
By considering the general case for $K$ and substituting Eq.~\eqref{eq:Nq_match_expectation} into Eq.~\eqref{eq:main_quantum_ENK}, we obtain
\begin{align}
    \sum_{z\in\mathcal{Z}_{Q\circ}}P(N_{Q\circ}(\Gamma_{Q})=z)^{K}\ge\cos\left[\frac{1}{2}\int_{0}^{\tau}\frac{\sqrt{\mathcal{B}(t)}}{t}dt\right]^{2K},
    \label{eq:PNq_quantum}
\end{align}
which derives Eq.~\eqref{eq:quantum_Renyi_bound} in the main text. 

\section{Quantum dynamical activity\label{sec:QDA}}

We briefly review the quantum dynamical activity,
a quantum generalization of the dynamical activity. 

First, we show how the classical dynamical activity defined in Eq.~\eqref{eq:dynamical_activity_def} can be expressed using the Lindblad equation. 
Using the jump operators $L_m$, the dynamical activity can be written as
\begin{align}
    \mathcal{A}_Q(\tau)\equiv\int_{0}^{\tau}\sum_{m}\mathrm{Tr}\left[L_{m}\rho(t)L_{m}^{\dagger}\right]dt.
    \label{eq:Atau_Lm}
\end{align}
In the classical limit, this expression reduces to the classical dynamical activity given in Eq.~\eqref{eq:Atau_def}.

Next, we consider a quantum generalization of the dynamical activity.
The quantum dynamical activity plays a central role in trade-off relations \cite{Hasegawa:2020:QTURPRL,Hasegawa:2023:BulkBoundaryBoundNC,
Nakajima:2023:SLD,Nishiyama:2024:ExactQDAPRE,Nishiyama:2024:OpenQuantumRURJPA,Yunoki:2025:FeedbackQTUR}. 
The exact expression for $\mathcal{B}(t)$ in the quantum dynamical activity was obtained in Refs.~\cite{Nakajima:2023:SLD,Nishiyama:2024:ExactQDAPRE}.
Specifically, Ref.~\cite{Nishiyama:2024:ExactQDAPRE} showed the following relation:
\begin{align}
    \mathcal{B}(\tau)=\mathcal{A}_{Q}(\tau)+\mathcal{C}(\tau),
    \label{eq:QDA_exact_solution}
\end{align}
where $\mathcal{C}(\tau)$ is the contribution from coherent dynamics,
\begin{align}
\mathcal{C}(\tau)&\equiv8\int_{0}^{\tau}ds_{1}\int_{0}^{s_{1}}ds_{2}\mathrm{Re}\left[\mathrm{Tr}\{H_{\mathrm{eff}}^{\dagger}\check{H}(s_{1}-s_{2})\rho(s_{2})\}\right]\nonumber\\&-4\left(\int_{0}^{\tau}ds\,\mathrm{Tr}\left[H\rho(s)\right]\right)^{2}.
    \label{eq:QDA_Bq_def}
\end{align}
Here, $\check{H}(t) \equiv e^{\mathcal{L}^{\dagger} t} H$ is the Hamiltonian $H$ in the Heisenberg picture,
$H_{\mathrm{eff}}=H-\frac{i}{2} \sum_m L_m^{\dagger} L_m$ is the effective Hamiltonian, 
and $\mathcal{L}^\dagger$ is the adjoint Liouvillian,
\begin{align}
    \mathcal{L}^{\dagger}\mathcal{O}\equiv i\left[H,\mathcal{O}\right]+\sum_{m=1}^{\mathfrak{N}}\mathcal{D}^{\dagger}[L_{m}]\mathcal{O},
    \label{eq:mathcalL_dagger_def}
\end{align}
with $\mathcal{D}^\dagger$ the adjoint dissipator,
\begin{align}
    \mathcal{D}^{\dagger}[L]\mathcal{O}\equiv L^{\dagger}\mathcal{O}L-\frac{1}{2}\{L^{\dagger}L,\mathcal{O}\},
    \label{eq:mathcalD_dagger_def}
\end{align}
for an operator $\mathcal{O}$. 

As discussed above, the classical dynamical activity quantifies how frequently the system undergoes jumps and can be obtained from jump statistics. 
The quantum dynamical activity extends this notion to quantum Markov processes. 
In such processes, the system state can change even in the absence of jumps because of coherent evolution generated by $H$, and $\mathcal{C}(\tau)$ in Eq.~\eqref{eq:QDA_exact_solution} captures this purely coherent contribution. 
In the classical limit $H=0$, we have $\mathcal{C}(\tau)=0$ and thus $\mathcal{B}(\tau)=\mathcal{A}(\tau)$.

Under steady-state conditions, the classical dynamical activity grows linearly in time, $\mathcal{A}(\tau) = \mathfrak{a}_\mathrm{ss}\tau$. 
By contrast, the quantum dynamical activity can exhibit superlinear growth over a finite time interval, even in the steady state, due to coherent dynamics \cite{Nishiyama:2024:ExactQDAPRE}. 
For large times, however, $\mathcal{B}(\tau)$ again becomes linear in $\tau$: in Ref.~\cite{Hasegawa:2020:QTURPRL}, the asymptotic form of $\mathcal{B}(\tau)$ for $\tau \to \infty$ was derived.

\begin{acknowledgments}

This work was supported by JSPS KAKENHI Grant Numbers JP23K24915 and JP24K03008.

\end{acknowledgments}

\end{document}